\documentclass{aastex631}
\usepackage{graphicx}

\shorttitle{Dust Buried Sources in NGC~4449}
\shortauthors{Calzetti et al.}

\graphicspath{{./}{figures/}}

\begin{document}

\title{Dust Buried Compact Sources in the Dwarf Galaxy NGC~4449}

\author[0000-0002-5189-8004]{Daniela Calzetti}
\affiliation{Department of Astronomy, University of Massachusetts Amherst, 710 North Pleasant Street, Amherst, MA 01003, USA}

\author[0000-0002-1000-6081]{Sean T. Linden}
\affiliation{Department of Astronomy, University of Massachusetts Amherst, 710 North Pleasant Street, Amherst, MA 01003, USA}

\author{Timothy McQuaid}
\affiliation{Department of Astronomy, University of Massachusetts Amherst, 710 North Pleasant Street, Amherst, MA 01003, USA}

\author[0000-0003-1427-2456]{Matteo Messa}
\affiliation{Department of Astronomy, Universit\'e de Gen\`eve, 24 rue du G\'en\'eral-Dufour, 1211 Gen\`eve 4, Switzerland }
\affiliation{Department of Astronomy, Stockholm University, Stockholm, Sweden}

\author[0000-0001-7673-2257]{Zhiyuan Ji}
\affiliation{Department of Astronomy, University of Arizona, Tucson, AZ, USA }

\author[0000-0003-3893-854X]{Mark R. Krumholz}
\affiliation{Research School of Astronomy and Astrophysics, Australian National University, Camberra, Australia}

\author[0000-0002-8192-8091]{Angela Adamo}
\affiliation{Department of Astronomy, Stockholm University, Stockholm, Sweden}

\author[0000-0002-1723-6330]{Bruce Elmegreen}
\affiliation{IBM Research Division, T.J. Watson Research Center, Yorktown Heights, NY, USA}

\author[0000-0002-3247-5321]{Kathryn Grasha}
\affiliation{Research School of Astronomy and Astrophysics, Australian National University, Camberra, Australia}

\author[0000-0001-8348-2671]{Kelsey E. Johnson}
\affiliation{Department of Astronomy, University of Virginia, Charlottesville, VA, USA}

\author[0000-0003-2954-7643]{Elena Sabbi}
\affiliation{Space Telescope  Science Institute, 3700 San Martin Drive, Baltimore, MD 21218, USA}

\author[0000-0002-0806-168X]{Linda J. Smith}
\affiliation{Space Telescope  Science Institute, 3700 San Martin Drive, Baltimore, MD 21218, USA}

\author{Varun Bajaj}
\affiliation{Space Telescope  Science Institute, 3700 San Martin Drive, Baltimore, MD 21218, USA}

\begin{abstract}
Multi--wavelength images from the {\sl Hubble Space Telescope} covering the wavelength range 0.27--1.6~$\mu$m show that the central area of the nearby dwarf galaxy NGC\,4449 contains several tens of compact sources that are emitting in the hydrogen recombination line Pa$\beta$ (1.2818~$\mu$m) but are only marginally detected in H$\alpha$ (0.6563~$\mu$m) and undetected at wavelengths $\lambda\le$0.55~$\mu$m. An analysis of the spectral energy distributions (SEDs) of these sources indicates that they are likely relatively young star clusters heavily attenuated by dust. The selection function used to identify the sources prevents meaningful statistical analyses of their age, mass, and dust extinction distributions. However, these cluster candidates have ages $\sim$5--6~Myr and A$_V>$6~mag, according to their SED fits, and are extremely compact, with typical deconvolved radii of 1~pc. The dusty clusters are located at the periphery of dark clouds within the galaxy and appear to be partially embedded.  Density and pressure considerations indicate that the HII regions surrounding these clusters may be stalled, and that pre--supernova feedback has not been able to clear the clusters of their natal cocoons. These findings are in potential tension with existing models that regulate star formation with pre--supernova feedback, since pre--supernova feedback acts on short timescales, $\lesssim$4~Myr, for a standard Stellar Initial Mass function. The existence of a population of dusty star clusters with ages $>$4~Myr, if confirmed by future observations, paints a more complex picture for the role of stellar feedback in controlling star formation.
 \end{abstract}


\section{Introduction} \label{sec:intro}
Star formation is the result of contrasting mechanisms acting on the gas in galaxies. Gravity and cooling collapses the gas producing high-density cores, where stars form; feedback from these stars disperses the natal cores, and injects energy and matter  into the interstellar (ISM) and circumgalactic medium, driving turbulence and maintaining a multiphase ISM \citep{Hennebelle+2011, Hopkins+2012, Dobbs+2015, Goldbaum+2016}. The feedback from stars --  outflows, radiation, winds, and supernovae (SNe)  -- regulates star formation either at the local level, by acting directly on the clouds to stop star formation \citep[e.g.,][]{Krumholz+2005, Dobbs+2011, Federrath2015, Krumholz+2019, Hopkins+2014, Grisdale+2017, Grudic+2018}, or at the global level, by maintaining galaxies in a "pseudo-equilibrium" state, thus setting the collapse rate on kpc scales \citep[e.g.,][]{Ostriker+2010, Ostriker+2022}. Once the fragmentation of a gas cloud begins, star formation is expected to be a fast process, taking less than a few Myr from the appearance of the first proto--stars to the disruption of the molecular cloud \citep[e.g.,][]{Grudic+2022}. The low efficiency of star formation, at the level of a few percent when integrated over large regions (giant molecular clouds, galaxies, etc.), is a direct consequence of stellar feedback from young, massive stellar populations \citep{Ostriker+2010, Hopkins+2014, Peters+2017, Orr+2018, Ostriker+2022}. 

Photoionization, direct and indirect radiation pressure and stellar winds are all part of massive star feedback and are often bundled under the terminology `pre--supernova (pre--SN) feedback', since they have timescales of a few Myr \citep{Pellegrini+2011, Dale+2012, Krause+2013, Krumholz+2019}. Pre--SN feedback helps clear the medium surrounding massive stars before the first supernova explosion occurs at around $\sim$4~Myr \citep{Leitherer+2014}, and the effect of supernova explosions is to mainly inject energy into the ISM outside of the natal cloud, rather than affecting the star formation within the natal cloud itself \citep{Lucas+2020, Grudic+2022}. According to models, radiative feedback may be key for regulating star formation within galaxies \citep{Hopkins+2020, Bending+2022} and processes that act over timescales shorter than supernova explosions may be required to clear channels in the ISM  for the escape of ionizing photons from galaxies, as half of the ionizing photons are supplied within the first 3~Myr \citep{Ma+2020}. 

Several observational studies support the short timescales of pre--SN feedback as the main mode to regulate star formation. These studies mainly  concentrate on young star clusters, as these contain the majority, $\gtrsim$70\%, of massive stars \citep{Oey+2004} and are therefore primary sites for studying stellar feedback. Combining UV/optical photometry of star clusters with H$\alpha$ morphology of HII regions from the {\em Hubble Space Telescope (HST)}, several authors have derived clearing timescales $<$4--5~Myr, and as short as 2~Myr, in nearby galaxies \citep{Whitmore+2011, Hollyhead+2015, Hannon+2019, Hannon+2022}. However, these studies use UV and optical data, and are thus limited to the relatively dust--free components of star formation. The addition of CO data to trace molecular clouds and, in some cases, 24~$\mu$m imaging from the {\em Spitzer  Space Telescope} to trace the dust--enshrouded star formation has enabled the use of both frequency and positional analysis to derive timescales; \citet{Matthews+2018}, \citet{Grasha+2018}, \citet{Grasha+2019}, \citet{Kruijssen+2019}, \citet{Kim+2021} and \citet{Chevance+2022} use different techniques to conclude that timescales are short, only 3--5 Myr, and likely shorter than the timescale for supernova explosions. \citet{Corbelli+2017} use multi--wavelength data, including {\em Spitzer  Space Telescope}'s 24~$\mu$m imaging, of the Local Group galaxy M\,33 to conclude that the embedded phase of star  formation is short,  $\sim$2~Myr. The main limitation of these studies is the use of low resolution data. CO data usually subtend regions $>50$~pc in size, which suppresses the contrast at the small--scales of star clusters \citep[$\sim$3~pc,][]{Ryon+2017, Brown+2021}. Even in the favorable case of  the nearby galaxy M\,33,  the Spitzer MIPS/24~$\mu$m data subtends a spatial scale of 26~pc, much larger than a single star cluster. Spitzer imaging of the Magellanic Clouds provides spatial resolution $\approx$1~pc, thanks to the Clouds' small distances \citep{Meixner+2006, Gordon+2011}, suitable to investigate feedback. However, infrared studies of the Magellanic Clouds mostly concentrate on either individual sources or specific classes of objects \citep[e.g., Young Stellar Objects,][]{Whitney+2008, Sewilo+2013}. In one analysis of HII regions in the Clouds, \citet{Lawton+2010} conclude that UV, H$\alpha$, and IR light track each other. Radio observations have been used to detect free--free emission from the gas ionized by newly formed massive stars embedded in their natal clouds  \citep[e.g.,][]{Kobulnicky+1999, Johnson+2001, Johnson+2003, Johnson+2009, Turner+2004, Kepley+2014}, with some authors  indicating an embedded timescale of $\sim$1~Myr \citep{Johnson+2003}. However, radio detection limits are usually too shallow to enable sampling of complete populations of star clusters \citep[e.g.,][]{Reines+2008}. 

In a pilot study, \citet{Messa+2021} analyze HST NUV--to--nearIR images of the nearby star--forming galaxy NGC\,1313, isolating a population of star clusters that are dustier than those identified from NUV--optical images only. These authors find that  about 40\%--60\% of the young star clusters ($<$6~Myr in age) are missed in UV--optically--selected catalogs; in addition, they extend the timescale for emergence of the clusters from the natal cocoon from $\approx$2--3~Myr to $\sim$3--4~Myr, edging closer to the timescale of supernova explosions. The main limitation of \citet{Messa+2021}'s work is to still require that the sources are detected at visible wavelengths in order to constrain their ages and masses. 

More recently, \citet{Linden+2022} use near--IR imaging data from the {\em JWST} to infer that about 2/3 of the star clusters younger than 4~Myr are missed by standard UV--optical observations  in the relatively nearby Luminous Infrared Galaxy VV114. \citet{Kim+2022} uses JWST mid--IR imaging, combined with H$\alpha$ and CO imaging, of the nearby star--forming galaxy NGC~628 to infer that the embedded phase of star formation lasts about 5~Myr, during the first half of which dust obscuration is so high that the H$\alpha$ emission is not detectable. In a complementary fashion, \citet{Whitmore+2022} combine HST optical with JWST near-- and mid--IR medium and broad--band imaging of the galaxy NGC\,1365 to conclude that massive (M$\sim$10$^6$~M$_{\odot}$) star clusters in this galaxy remain completely or partially obscured for about 4$\pm$1~Myr. However, \citet{Linden+2022}, \citet{Kim+2022} and \citet{Whitmore+2022} do not have information on the infrared hydrogen recombination lines, which are key for constraining the ages of young star clusters when these are marginally detected or undetected at optical wavelengths. Clearly, more investigations are needed to obtain a census of the fraction of dust--embedded star formation in  galaxies and address its implications for the efficacy of pre--SN feedback. 

In this work, we attempt to mitigate earlier limitations with resolution and wavelength range by analyzing the compact source content of the nearby starburst dwarf galaxy NGC\,4449 (section~\ref{sec:galaxy}) with broad and narrow--band NUV--to--nearIR imaging data from the {\em HST}, covering the range 0.27--1.6~$\mu$m with angular resolution $\sim$0$^{\prime\prime}$.2. This angular resolution subtends a spatial scale of $\sim$4~pc at the distance of the galaxy, comparable to the size of star clusters \citep{Ryon+2017, Brown+2021}. We also move away from standard optical--centered detections and require our sources to be undetected in the V--band and bluer wavelengths, to attempt to secure truly dusty star clusters.

\section{The Dwarf Starburst Galaxy NGC\,4449} \label{sec:galaxy}

NGC\,4449 is a nearby  Magellanic irregular dwarf galaxy with $\sim$1/3 the stellar mass of the Large Magellanic Cloud (Table~\ref{table:galaxy}). Its star formation rate (SFR) is sufficiently high to place this galaxy about three times or more above the  Main Sequence of Star Formation for local galaxies \citep{Whitaker+2012, Cook+2014}, and thus qualify it as a starburst galaxy. The galaxy emits strongly at all wavelengths, including in the light of hydrogen recombination lines, in agreement with being rapidly forming stars at the present time.  Star formation is centrally concentrated, with 90\% located within 57\% of R$_{25}$, as derived from dust attenuation--corrected UV images \citep[e.g.,][]{Calzetti+2018}. For context, this corresponds to a SFR surface density, $\Sigma_{SFR}$=0.03~M$_{\odot}$~yr$^{-1}$~kpc$^{-2}$, about six times higher than the average in the Large Magellanic Cloud. The starburst in NGC\,4449 has been triggered by either a minor merger or the interaction with another galaxy \citep{Hunter+1998, Lelli+2014}. The metal content is $\sim$40\% of the solar value\footnote{We adopt a solar oxygen abundance of 12+Log(O/H)=8.69, \citet{Asplund+2009}.} \citep{Berg+2012}, with a modest gradient \citep[Table~\ref{table:galaxy},][]{Pilyugin+2015}. The slightly sub--solar metallicity is consistent with the galaxy having a modest dust content: its IR/UV ratio indicates that only 40\% of the light from young stars is absorbed by dust in this galaxy \citep{Hao+2011, Grasha+2013}.

\begin{deluxetable}{rrr}
\tablecolumns{3}
\tabletypesize{\footnotesize}
\tablecaption{Adopted Properties for NGC\,4449.\label{table:galaxy}}
\tablewidth{0pt}
\tablehead{
\colhead{Parameter (Units)} &  \colhead{Value} & \colhead{Reference\tablenotemark{a}} 
\\
}
\startdata
\hline
Morphology  & IBm & 1 \\
Distance (Mpc)  & 4.2         & 2      \\
Recession  velocity (km/s) & 207 & 3 \\
Inclination (degrees) & 68 & 4 \\
E(B-V)$_{MW}$\tablenotemark{b} & 0.017 & 5 \\
Stellar Mass (M$_{\odot}$)    & 1$\times$10$^9$    & 6      \\
SFR (M$_{\odot}$~yr$^{-1}$)\tablenotemark{c}  & 0.5      &  7       \\
12$+$Log(O/H)\tablenotemark{d}  & 8.26 & 8 \\
Metall. Gradient (dex/kpc)\tablenotemark{d}   & $-$0.055          & 9    \\
$[NII]/$H$\alpha$ & 0.11 & 8 \\
\hline
\enddata

\tablenotetext{a}{1 -- \citet{DeVac+1991}; 2 -- \citet[][from TRGB]{Tully+2013};  3 --  \citet{Schneider+1992}; 4 -- \citet{Hunter+2005}; 5 -- \citet{Schlafly+2011}; 6  -- \citet{Calzetti+2015a}; 7 -- \citet{Lee+2009};  8 -- \citet{Berg+2012};  9 -- \citet{Pilyugin+2015}.}
\tablenotetext{b}{Foreground Milky Way color excess.}
\tablenotetext{c}{Star formation rate from the dust attenuation--corrected ultraviolet luminosity.}
\tablenotetext{d}{Central oxygen abundance and abundance gradient, respectively. The central oxygen abundance has an uncertainty of $\pm$0.09 \citep{Berg+2012}.}
\tablecomments{Data and references obtained from NED, the NASA/IPAC Extragalactic Database.}
 \end{deluxetable}

\section{Imaging Data and Processing} \label{sec:data}

The {\em Hubble Space Telescope} (HST) observations used in this work were obtained from several programs, and cover the near--UV to near--IR range in 10 bands, listed in Table~\ref{table:images}. The images obtained with the WFC3/IR camera are part of the program GO--15330, and include two continuum band filters (F110W and F160W) and one 
narrow--band filter (F128N) centered on the Pa$\beta$(1.2818~$\mu$m) hydrogen recombination line emission. For each filter, the standard calibration pipeline CALWFC3 v. 3.5.2 was used to process individual frames into the final images, after correction for bias, dark, and flat--field. The UV and optical images were retrieved from the archive already processed through their respective instrumental pipelines. All images were aligned, mosaicked and resampled to the smallest pixel scale, 0$^{\prime\prime}$.04/pix, using DrizzlePac; alignment was performed on the Gaia DR2 reference frame. At the distance of NGC\,4449, 0$^{\prime\prime}$.04 subtends 0.81~pc. The final images are in units of e$^-$/s, and calibration to physical flux is performed by applying the image header keyword PHOTFLAM. Each filter's central wavelength and the limiting flux density reached in each image are listed in Table~\ref{table:limiting}; the limiting flux density is calculated from the standard deviation of the Eastern half of each image's field--of--view (FoV), where the galaxy's emission is fainter than the average. The limiting flux is determined partly by the exposure time  (i.e., the depth of the images) and partly by the density fluctuations in the galaxy's unresolved stellar field, since NGC~4449 is large enough to completely fill the FoV of the HST images. Within a small region, the limiting flux density can be higher or (slightly) lower than the average, depending on the local stellar density. The WFC3--IR  camera has the smallest FoV among the instruments used, and determines the spatial coverage adopted in this analysis; it corresponds to the central, most active, 2.8$\times$2.4~kpc$^2$ of the galaxy. This region encompasses 67\% of the total SFR of the galaxy.

 Emission line images are obtained by subtracting the stellar continuum from the narrow--band images. The stellar continuum for the F658N (H$\alpha+$[NII]) image is constructed from the interpolation between the F550M and the F814W, both tracers of stellar emission with only weak emission lines. The stellar continuum for the F128N (Pa$\beta$) image is  obtained from the interpolation between the F110W and the F160W. Since the F110W also contains the Pa$\beta$ line emission, the subtraction is performed iteratively; two iterations are sufficient for convergence to the final continuum--subtracted image. Both continuum--subtracted images are then multiplied by the respective filter widths (Table~\ref{table:limiting}) and corrected for the filter transmission curve at the galaxy's redshift (Table~\ref{table:galaxy}), in order to derive line fluxes. The optical line is further corrected for the [NII] contribution, using the value of [NII]/H$\alpha$ from Table~\ref{table:galaxy}. The final result is two emission--line images at H$\alpha$($\lambda$0.6563~$\mu$m) and Pa$\beta$($\lambda$1.2818~$\mu$m), respectively. A second near--IR line emission image is constructed at twice the pixel scale (0$^{\prime\prime}$.08/pix) of our default, since the Point Spread Function (PSF) of WFC3/IR images is about twice that of the optical images, 0$^{\prime\prime}$.19 versus 0$^{\prime\prime}$.08. This second Pa$\beta$ line emission image is used to exclude artificial Pa$\beta$ emission `sources' that may be produced by slight misalignments between the three near--IR images. 

\begin{deluxetable*}{llll}
\tablecaption{Imaging Data Sources\label{table:images}}
\tablewidth{0pt}
\tablehead{
\colhead{Instrument$^1$} & \colhead{Filters$^2$} & \colhead{Exposure Times$^3$} &
\colhead{Proposal ID$^4$}\\
}
\decimalcolnumbers
\startdata
WFC3/UVIS   &  F275W, F336W & 2480, 2360 & 13364\\
ACS/WFC   &  F435W, F550M, F555W, F658N, F814W & 7140, 1200, 4920, 2260, 4660 & 10522, 10585\\
WFC3/IR   &  F110W, F128N, F160W & 1000, 2600, 1700 & 15330 \\
\enddata
$^1$  WFC3/UVIS=Wide Field Camera 3 UV--Optical channel. ACS/WFC= Advanced Camera for Surveys Wide Field Channel. WFC3/IR=Wide Field Camera 3 Infrared channel. \\
$^2$ HST Filter names.\\
$^3$ The total exposure time in each filter in seconds.\\
$^4$ Identification of the GO program that obtained the images: GO--13364 (LEGUS, Legacy ExtraGalactic UV Survey), PI: Calzetti; GO--10522, PI: Calzetti; GO--10585, PI: Aloisi; GO--15330, PI: Calzetti.\\
\end{deluxetable*}

\begin{deluxetable*}{llrcrr}
\tablecaption{Imaging Data Characteristics\label{table:limiting}}
\tablewidth{0pt}
\tablehead{
\colhead{HST Filter$^1$} & \colhead{Pivot $\lambda$ $^2$} & FWHM$^2$ & \colhead{Standard Filter$^2$} & \colhead{Limiting Flux Density$^3$} &  \colhead{$\kappa$($\lambda$)$^4$} \\
\colhead{} & \colhead{$\mu$m} & \colhead{$\mu$m} & \colhead{} & \colhead{erg~s$^{-1}$~cm$^{-2}$~\AA$^{-1}~pix$$^{-1}$ } & \colhead{} \\
}
\decimalcolnumbers
\startdata
F275W & 0.2710 & 0.04053& NUV & 1.51$\times$10$^{-20}$ & 6.285\\
F336W & 0.3355 & 0.05116 & U & 8.87$\times$10$^{-21}$ & 5.068 \\
F435W & 0.4329 & 0.06911& B & 9.17$\times$10$^{-21}$ & 4.186\\
F550M & 0.5581 & 0.03845 & Medium V & 1.13$\times$10$^{-20}$ & 3.045\\
F555W & 0.5360 & 0.08478& V & 9.23$\times$10$^{-21}$ & 3.191\\
F658N  & 0.6584 & 0.00875& H$\alpha$+[NII] & 1.47$\times$10$^{-20}$ & 2.525\\
F814W & 0.8048 & 0.15416 & I & 6.26$\times$10$^{-21}$ & 1.831\\
F110W & 1.1534 & 0.44300& J & 1.91$\times$10$^{-21}$ & 0.995\\
F128N & 1.2832 & 0.01590 & Pa$\beta$ & 1.82$\times$10$^{-21}$ & 0.838\\
F160W & 1.5369 & 0.26830 & H & 1.81$\times$10$^{-21}$ & 0.627\\
\enddata
$^1$  WFC3/UVIS=Wide Field Camera 3 UV--Optical channel. ACS/WFC= Advanced Camera for Surveys Wide Field Channel. WFC3/IR=Wide Field Camera 3 Infrared channel. \\
$^2$ Filter names, pivot wavelength, Full Width at Half Maximum (FWHM) (from the  STScI Instrument Handbooks) and closest Standard Photometric System filter \citep{Bessel2005}. For the narrow--band filters, the main lines targeted are: H$\alpha$(0.6563~$\mu$m) and the [NII] (0.6548,0.6584~$\mu$m) doublet in F658N and Pa$\beta$(1.2818~$\mu$m) in F128N.\\
$^3$ The reported limiting flux densities are 1~$\sigma$ values averaged across the Eastern 1/2 of the FoV of the images. The pixel size is 0$^{\prime\prime}$.04. \\
$^4$ The values of the extinction curve: $\kappa$($\lambda$)=A(V)/E(B--V) adopted in this work \citep[from][]{Fitzpatrick+2019} to correct fluxes for the foreground Milky Way dust extinction.\\
\end{deluxetable*}

\section{Source Selection} \label{sec:selection}

Our goal is to isolate sources that are emitting in hydrogen recombination line, thus are possibly young star clusters, and are heavily attenuated by dust. Sources that are detected in Pa$\beta$, but are undetected in H$\alpha$ would qualify as heavily attenuated, if they exist. We thus elect to visually inspect our images, searching for compact sources that are detected in F110W, Pa$\beta$, and F160W, are weakly detected in F814W, are marginally detected (S/N$\lesssim$3) in F658N, generally undetected or barely detected in the H$\alpha$ line,  and are undetected in any filter bluer than F658N. When evaluating the detection level of the line emission from the sources, we consider that it may be nested amid diffuse emission. 

Qualitatively, we already know what kind of sources we are likely to identify. For a line emitting source with about 1/2 solar metallicity, H$\alpha$/Pa$\beta$=17.6 \citep{Osterbrock+2006, Calzetti+2007}. If we require that Pa$\beta$ is detected at least at the 3~$\sigma$ level and H$\alpha$  is undetected (i.e., $\lesssim$1~$\sigma$), we constrain our sources to have at least a color excess E(B$-$V)$\gtrsim$2.1 mag (A(V)$\gtrsim$6.5~mag). These values translate in a minimum suppression of a factor $\sim$135 for H$\alpha$ and $\sim$5 for Pa$\beta$. The actual minimum value of E(B$-$V) will be higher than 2.1~mag as our estimate does not include the contribution of the stellar continuum to the narrow--band filters. The calculation assumes a foreground dust geometry and takes into account that the F658N image has twice the angular resolution of the F128N image and is 8 times less deep (Table~\ref{table:limiting}).

We  use the Pa$\beta$ line emission images at both pixel scales: 0$^{\prime\prime}$.08/pix and 0$^{\prime\prime}$.04/pix to ensure  that our detections do not suffer from biases due to small misalignments among the images which could create spurious `emission' sources. We further inspect all candidate sources for potential artifacts, such as saturation in F110W and F160W, which would create an artificially low stellar continuum image under the F128N filter resulting in `emission--line sources'. We test the effects of continuum under--subtraction, which would also artificially create `emission--line sources' in the narrow--band filter, by multiplying the two IR broad band images by increasing factors up  to  $\sim$15\%; this value represents a hard upper limit to the combined uncertainties of the continuum interpolation and the absolute photometric calibration of the WFC3 bands\footnote{https://hst-docs.stsci.edu/wfc3dhb/chapter-7-wfc3-ir-sources-of-error/7-11-ir-photometry-errors}. The effect of  this  operation on the Pa$\beta$ image is to remove the diffuse emission, creating obvious over--subtracted images, and depress the compact source line emission, but without removing it. Thus, both saturation and under--subtraction are excluded as problems for our sources. We finally exclude any source that has another source within an aperture of 5~pixels  (0$^{\prime\prime}$.20) radius in the F814W and shorter wavelength images; this is to ensure that no contamination from neighboring sources (including their wings) affects the photometry in the lower resolution WFC3/IR images, which have PSF FWHM=0$^{\prime\prime}$.19 ($\sim$5~pix). 

\begin{figure}
\plotone{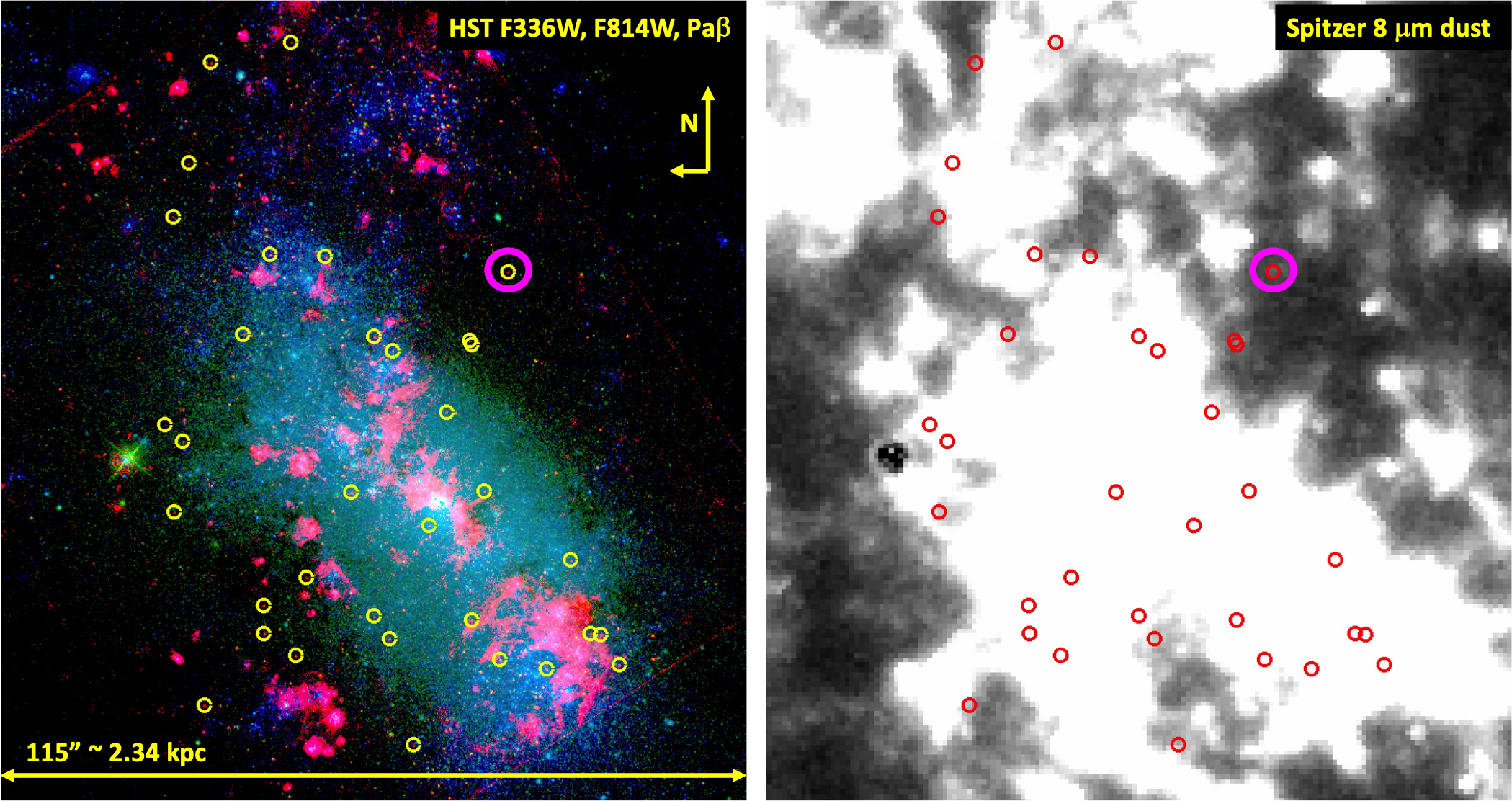}
\caption{Location of the 34 Pa$\beta$ emitting sources shown on a three--color composite (left; blue=F336W, green=F814W, red=Pa$\beta$ emission) and on a 8~$\mu$m dust image (right) of NGC\,4449. The three--color composite uses the HST images discussed in this paper; the 8~$\mu$m dust image is obtained from the Spitzer IRAC 8~$\mu$m image after removal of the stellar continuum by subtracting the IRAC 3.6~$\mu$m image \citep[e.g.][]{Calzetti+2018}. The location of the sources is marked by yellow (left) and red (right) circles, respectively. All identified sources are in correspondence of strong dust emission from the galaxy. The magenta circle identifies the source shown in detail in Figure~\ref{fig:cluster}. North  is up, East is left.}
 \label{fig:galaxy}
\end{figure}

\begin{figure}
\plotone{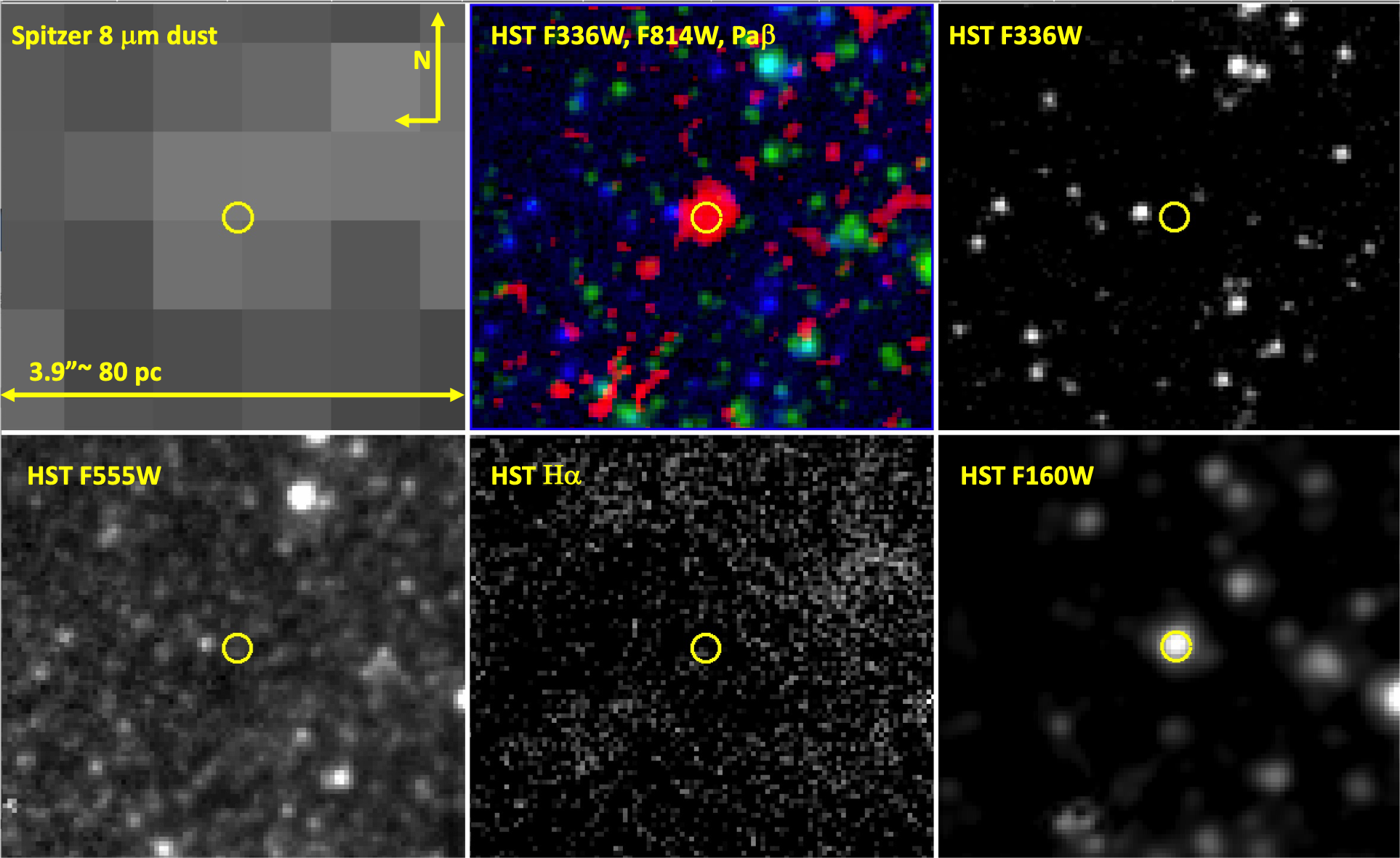}
\caption{The location of source \# 29 indicated by a yellow circle and shown at a range of wavelengths. The yellow circle has a radius of 3~pix (0$^{\prime\prime}$.12), which is the size of our photometric apertures. This is the cluster marked with a magenta circle in Figure~\ref{fig:galaxy}. Despite being located in a relatively dust--faint area of the galaxy, its 8~$\mu$m dust emission is still detected at 9~$\sigma$. The cut--outs show (from top--left to bottom--right): the Spitzer 8~$\mu$m dust emission, the HST three--color composite as in the previous figure, and single--band HST emission in the light of F336W (U), F555W (V), H$\alpha$, and F160W (H). Like all sources considered in the present analysis, this source is only detected in the I--band (F814W) and at longer wavelengths and marginally detected in F658N, but is undetected at shorter wavelengths. In particular, while the Pa$\beta$ emission is strong, the H$\alpha$ emission is generally not or only weakly detected above the diffuse background. 
\label{fig:cluster}}
\end{figure}

\begin{figure}
\plotone{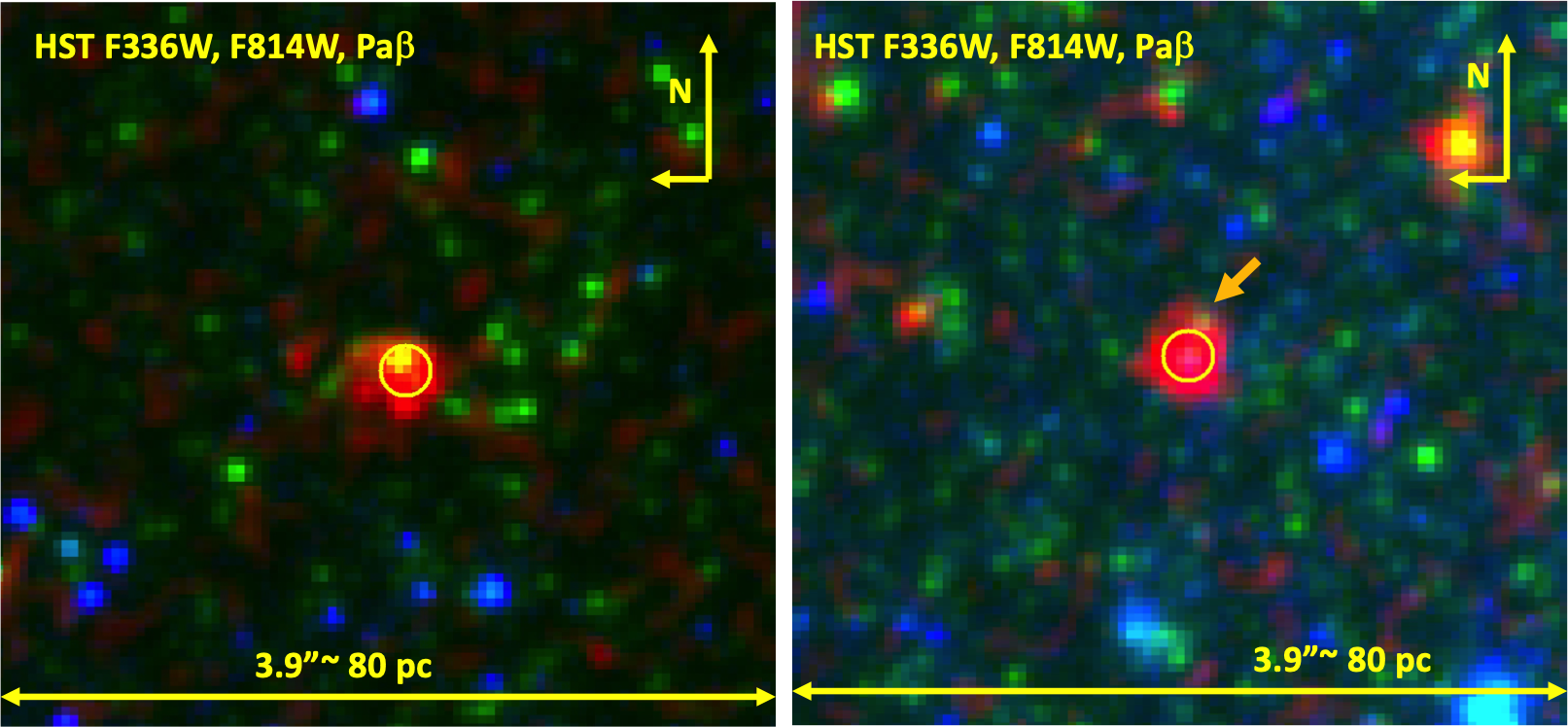}
\caption{Two examples of discarded sources, marked by a yellow circle. Like in Figure~\ref{fig:cluster}, the yellow circle has a radius of 3~pix (0$^{\prime\prime}$.12). The source to the left was discarded because the  continuum source is slightly misaligned with  the center of the line emission, raising the suspicion that the observed line emission may be the result of under--subtraction of the underlying stellar continuum due to potential offsets between the narrow--band and  broad--band filters. The source to the right was discarded because of the  presence of an adjacent optically--emitting source (top--right of the yellow circle, indicated by an orange arrow), which may affect  photometry in the IR bands. The HST three--color composite is as in the previous figure.  
\label{fig:discarded}}
\end{figure}

After all selections above are implemented, the original sample of $\sim$60 candidates is reduced to 34 sources, listed in Table~\ref{tab:photometry}. The location of the 34 sources is shown in Figure~\ref{fig:galaxy}. An example of the sources in our sample is shown in Figure~\ref{fig:cluster}, while two examples of `discarded' sources are shown in  Figure~\ref{fig:discarded}. The breakdown of reasons for discarding sources is as follows: about 2/3 are rejected because the Pa$\beta$ line emission appears misaligned with the  continuum images, which may potentially result in under--subtraction of the narrow--band F128N image\footnote{These sources could have misaligned line--continuum simply because the gas is physically displaced from the centroid of the stellar continuum, and thus be true line--emitting sources. However, the low resolution of the WFC3/IR images does not enable us to assess this scenario, and the sources are conservatively discarded.}; of the remaining 1/3 discarded sources, about 1/2 are rejected because detected in F555W and 1/2 because of nearby contaminating sources. 

A comparison with the distribution of the dust emission traced by the Spitzer Space Telescope's 8~$\mu$m image shows that all 34 retained sources are in areas of the galaxy where  dust emission is present. The low resolution of the Spitzer images does not allow  us to obtain deeper insights, except to estimate that even  the most `dust faint' among our sources resides in an area with 8~$\mu$m dust emission detected at least to the 9~$\sigma$ level (Figure~\ref{fig:cluster}). The 8~$\mu$m dust image for NGC\,4449 is derived from the IRAC 8~$\mu$m mosaic, after subtraction of the stellar continuum using the IRAC 3.6~$\mu$m image \citep[see description and details in][]{Calzetti+2018}. The coincidence of our sources with  8~$\mu$m dust--emitting regions in the galaxy further reinforces that our sources are unlikely to be image artifacts. The identified sources are extremely compact with several unresolved sources, as measured in the F814W filter; the most frequent (deconvolved) size is $\approx$1~pc ($\approx$0$^{\prime\prime}$.05), with a range  $<$0.5--5~pc. The sources are consistent with point--sources in the near--IR bands, including in the line emission, although this is to be expected as the near--IR images have lower angular resolution than the I--band image. If the sources are star clusters, they are extremely compact, similar to  the result by \citet{Messa+2021} for NGC\,1313. 

\section{Photometry} \label{sec:photometry}

Photometry is performed in circular apertures on all sources in all bands. The apertures are chosen with 3~pixels radius (0$^{\prime\prime}$.12=2.44~pc) on the plane of the sky to ensure that all photometry, and especially the upper limits (all bands shortwards of F658N), are not affected by the presence of neighboring contaminants, like, e.g., in the case of the discarded source  in Figure~\ref{fig:discarded}, right. The photometry is background--subtracted using a background annulus 5 pixels in radius and 2 pixels in thickness. Aperture corrections are applied to the measurements to account for the flux outside of the 3~pixel aperture. As the sources are consistent with point or extremely compact sources (see previous section), we use isolated stellar sources in the images to measure the curves of growth and the aperture corrections. We derive the following aperture corrections: 1.60$\pm$0.04 for F814W, 2.54$\pm$0.05 for F110W, 2.65$\pm$0.05 for F128N, and 2.91$\pm$0.04 for F160W. We apply the F814W aperture corrections also to the F658N measurements and the upper limits at bluer bands, based on the results of \citet{Messa+2021} who determine that differences in aperture corrections between HST UV and I band are $\lesssim$10\%. Finally, all photometry is corrected for the MW foreground extinction, E(B--V)=0.017 (Table~\ref{table:galaxy}). Uncertainties in the photometry are the geometric combination of photon noise, uncertainty in the aperture correction and the standard deviation of the background measurement (which also includes instrumental uncertainties). The final photometry is listed in Table~\ref{tab:photometry}, together with the luminosity and equivalent width  (EW) of Pa$\beta$. 

Our 2.44~pc (3~pix) aperture radius on the plane of the sky corresponds to an ellipse of 2.44$\times$6.51~pc$^2$ on the plane of the galaxy, based on its inclination
 (Table~\ref{table:galaxy}); this is equivalent to a circular aperture (by area) of 3.5~pc radius. \citet{Ryon+2017} finds that the peak of the distribution of effective radii of  star clusters in two galaxies of the LEGUS sample, which is a sample of  galaxies within $\sim$12~Mpc \citep{Calzetti+2015a}, is at 2--3~pc. \citet{Brown+2021} finds, for the full LEGUS sample of galaxies, a shallow correlation between cluster effective radii and stellar mass: clusters with mass $\sim$10$^4$~M$_{\odot}$ have effective radii $\sim$2.6~pc, which rescales according to the relation $R_{eff}\propto M^{0.24}$ for different masses. Thus, if our sources are stellar clusters, the photometric aperture we have chosen is a reasonable compromise between capturing the clusters' emission and minimizing contamination from neighboring sources.

\section{Models and Fitting Approach} \label{sec:models}

We compare the photometry of the 34 sources measured above (section~\ref{sec:photometry}) with models, in order to derive physical properties: age, mass, and extinction,
 under the assumption that presence of the hydrogen recombination line emission is likely tracking young stellar clusters. We generate synthetic photometry from the Yggdrasil SED models \citep{Zackrisson+2011} combined with dust attenuation/extinction recipes, following the same procedure outlined in \citet{Calzetti+2015b}, \citet{Adamo+2017}, \citet{Messa+2021} and \citet{Calzetti+2021} and briefly summarized here. 
 
Yggdrasil uses the Starburst99 \citep{Leitherer+1999} spectral synthesis models as an input for CLOUDY \citep{Ferland+2013}, to produce single stellar population (SSP) models that include, in addition to stellar and nebular continuum (a standard output of Starburst99), also nebular emission lines. The SSP models from Starburst99 are generated using instantaneous star formation, with a \citet{Kroupa2001} IMF in the range 0.1--120~M$_{\odot}$ and metallicity Z=0.008 ($\sim$40\% solar), which is the closest value to the measured oxygen abundance of NGC\,4449 and for which models are available. The models are produced using both the Padova with AGB treatment and the Geneva tracks \citep{Meynet+1994, Girardi+2000, Vazquez+2005}. From the Yggdrasil suite, we adopt the models with a 50\% covering factor for the ionized gas, meaning that only 50\% of the Lyman continuum photons produced by the star cluster ionize the gas. Although we do not have a handle on the actual gas covering factor of our sources, it is reasonable to expect that not all ionizing photons are available to ionize the surrounding gas: at the large extinctions of our sources, a significant portion of the ionizing photons is likely to be absorbed directly by the dust in which the sources are embedded \citep{Dopita+2003}. Models are generated with ages between 1~Myr and 14~Gyr; however, we are interested in the youngest (ionizing) ages, $<$10~Myr, for which the models are generated in 1~Myr steps. We determine a posteriori that the sources under consideration are massive, with median M$\sim$5,600~M$_{\odot}$ and a minimum mass of $\sim$3,000~M$_{\odot}$. This, coupled with the fact that all our sources are detected in hydrogen recombination line emission, implies that we expect minimal impact from stochastic sampling of the IMF \citep{Cervino+2002}, and use deterministic models, i.e., full sampling of the stellar IMF, for the derivation of the physical parameters. 

The SSP SEDs are then attenuated with: a starburst attenuation curve \citep{Calzetti+2000} and an LMC  extinction curve \citep[as parametrized by][]{Fitzpatrick1999}. We adopt a foreground dust geometry \citep{Calzetti2001} of the form:
\begin{equation}
F(\lambda)_{final} = F(\lambda)_{model} 10^{[-0.4 E(B-V) \kappa(\lambda)]},
\end{equation}
where E(B$-$V) is the color excess and $\kappa(\lambda)$ is the attenuation/extinction curve. No other extinction curve beyond the LMC is considered here, because our sources are undetected at V and bluewards, and functional shape differences among extinction curves are only found below the V band \citep{Gordon+2003}. For the LMC extinction curve, both cases of equal and differential attenuation for the nebular gas and stellar continuum are considered; for the differential attenuation, we assume that the stellar continuum is subject to about half the attenuation of the nebular gas \citep{Calzetti+1994,Kreckel+2013}. For the starburst attenuation curve, the dust geometry, including the differential attenuation, is `built--in' into the functional form of the curve.  In summary, we generate models with three types of extinction/attenuation: LMC, LMC with differential attenuation between ionized gas and stellar emission, and starburst. We generate the models in the color excess range E(B$-$V)=0--5~mag, with step 0.02. 

We only consider foreground dust attenuation/extinction because more complex geometries, including mixed dust/star/gas geometries, maximize total attenuation while minimizing differential attenuation \citep{Gordon+1997, Calzetti2001}. For instance, in the extreme case of homogeneously  mixed dust and stars/gas with large dust column density, the net effect is to drastically dim the source while maintaining a blue overall SED \citep{Calzetti+1994, Calzetti+2015b}. In our case, we are trying to select geometries that maximize differential attenuation, so to abate the emission in the blue bands while keeping the I and near--IR bands above their respective detection thresholds. 

The dust--attenuated model SEDs are convolved with the transmission curve of the filters plus the HST optics to produce synthetic photometry for all six model combinations (2 tracks and 3 extinction/attenuation recipes). The synthetic photometry is then compared with the measured photometry using $\chi^2$--minimization, taking into account the measurement uncertainties, to obtain the distribution of solutions and the reduced 
$\chi^2$ value for each source. We finally plot the distribution of solutions within the 90\% significance level  for the appropriate number of degrees of freedom, and select the 
best values and the 68\% (1~$\sigma$) uncertainty for the parameters of each region based on the shape of the reduced $\chi^2$ probability distribution. 

The presence of emission in the Pa$\beta$ line already enables us to break some degeneracies in the fits, since our sources are likely to have young ages. Thus, we do not expect the usual age--extinction degeneracy between `dusty and young'  and `dust--free and old' often found in SED fits \citep{Whitmore+2020}. 

Although we measure photometry in 10 separate bands, we have upper limits for the 5 bluest ones. The reddest upper limit, the luminosity density in F550M, provides the bluest useful constraint to the SED fits, meaning that we are effectively using 6 datapoints for the fits, thus we have 3 degrees of freedom. Below the F550M upper limit, we do not consider deviations between the remaining upper limits and the fits as being meaningful. We include the narrow--band F658N and F128N filters in the fits, although gas emission can have a different spatial distribution from that of the stars \citep{Calzetti+2015b}. We do  this for two reasons: (1) we have a limited number of measurements for our SEDs (excluding upper limits), and (2)  our sources are point--like at all bands where they can  be measured, suggesting that the sources are confined by the surrounding medium and the ionized gas  is co--spatial with the stars. 

\section{Results} \label{sec:results}

\begin{figure}
\plottwo{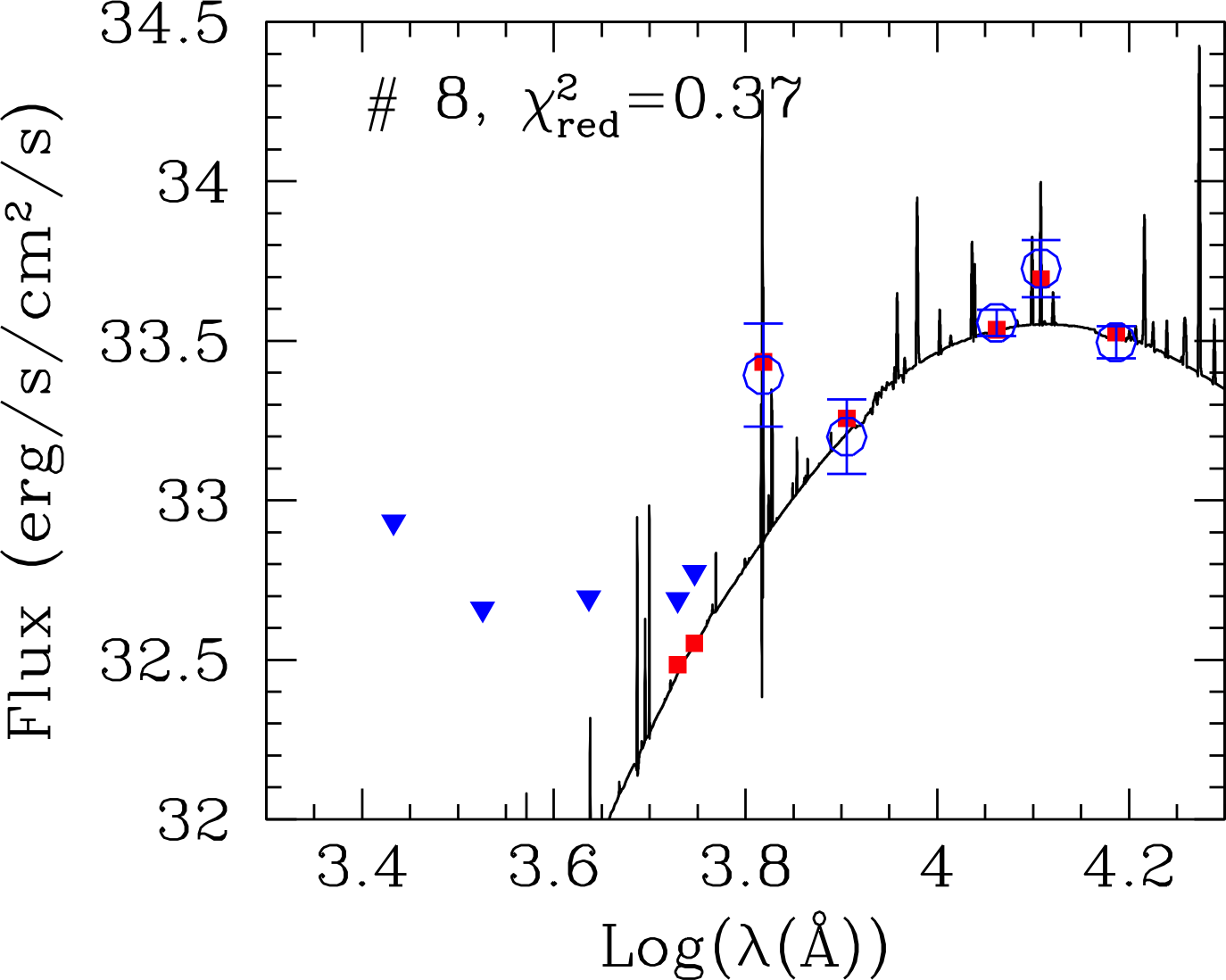}{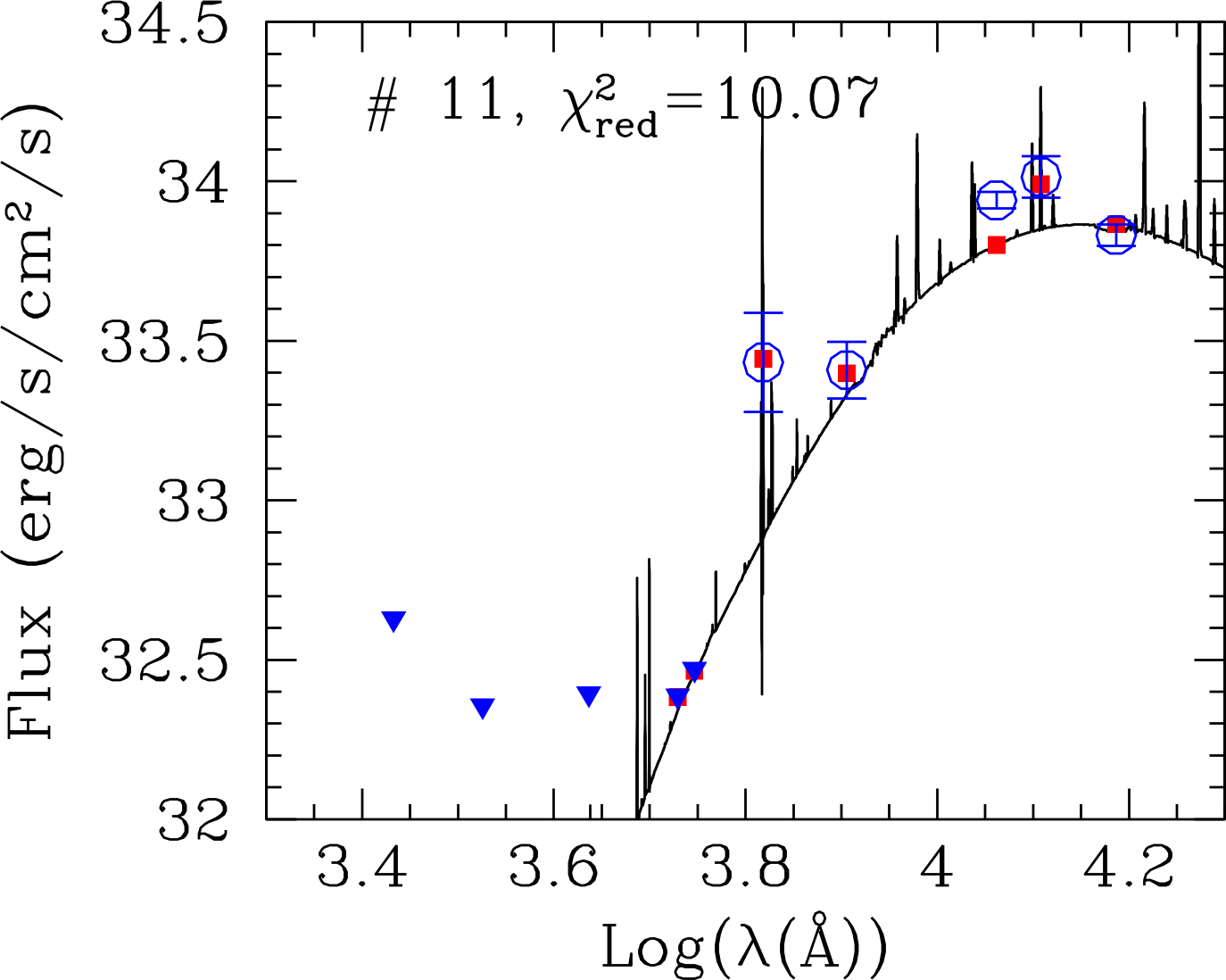}
\caption{The measured and best--fit photometry and model spectra for two sources, showing two very different cases: an excellent goodness--of--fit ($\chi^2_{red}<1$) and a low--significance one ($\chi^2_{red}>10$). The measurements (circles with 1~$\sigma$ error bars) and upper limits (downward triangles) are shown in blue. Best--fits photometry is shown as red squares, and best--fit spectrum is shown as a black line. The source ID is listed in each panel, and corresponds to the IDs listed in both Tables~\ref{tab:photometry} and \ref{tab:source_properties}.} 
\label{fig:SEDs}
\end{figure}

\begin{figure}
\plottwo{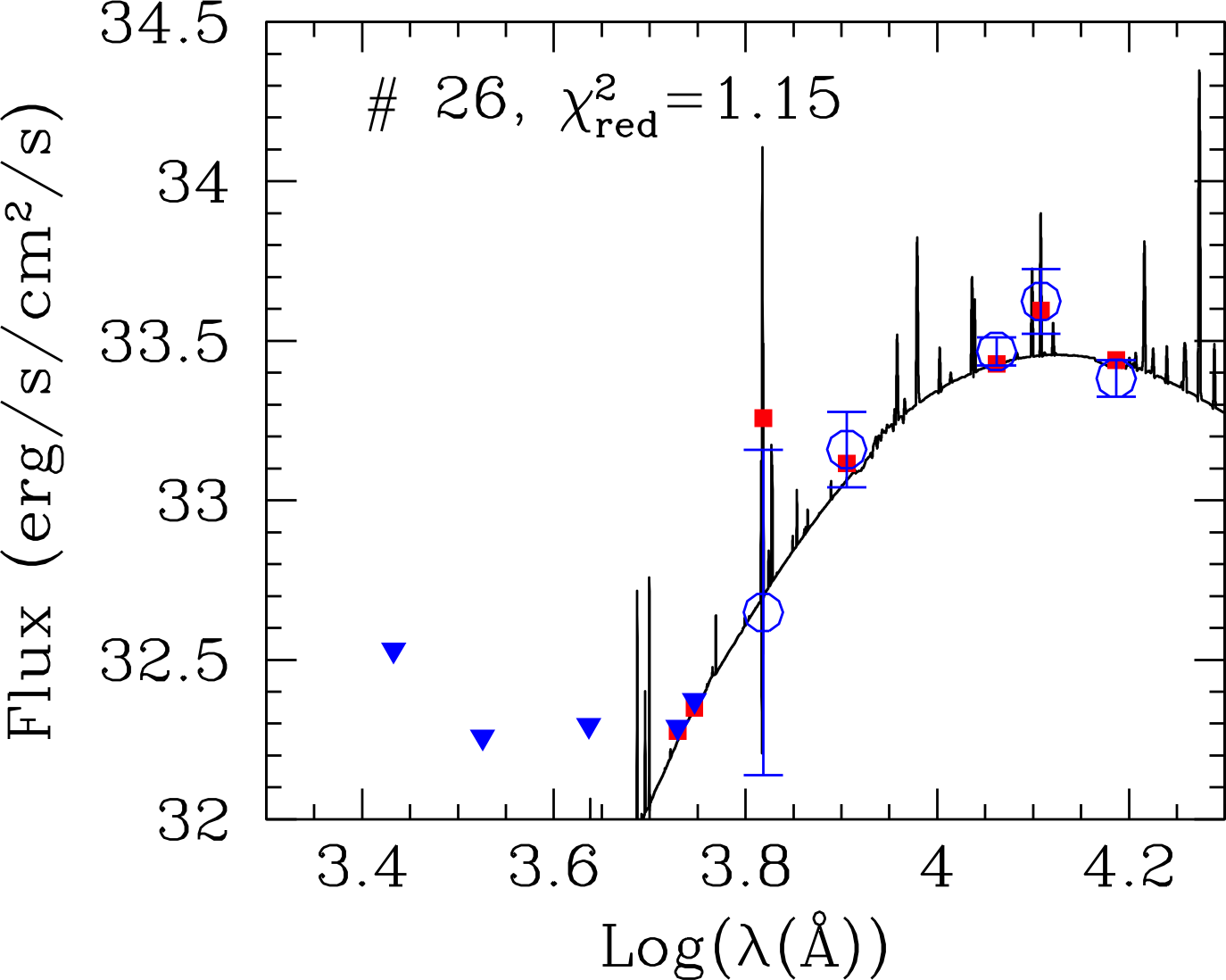}{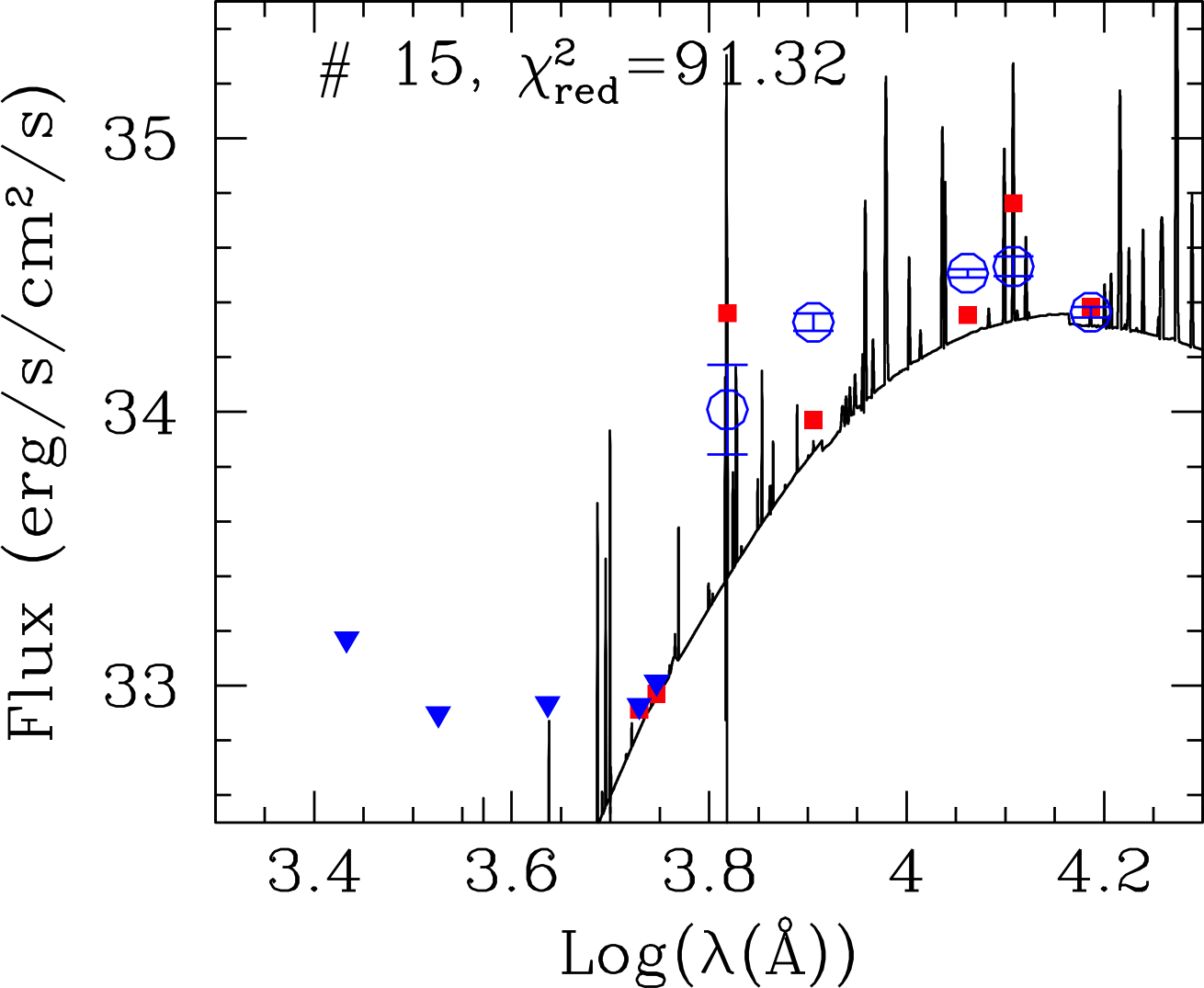}
\caption{Two additional examples of observed photometry and best fit for our sources, including the case of the worse $\chi^2_{red}$ in our sample (right panel). Symbols and lines are as in Figure~\ref{fig:SEDs}.} 
\label{fig:SEDs_2}
\end{figure}

\begin{figure}
\plottwo{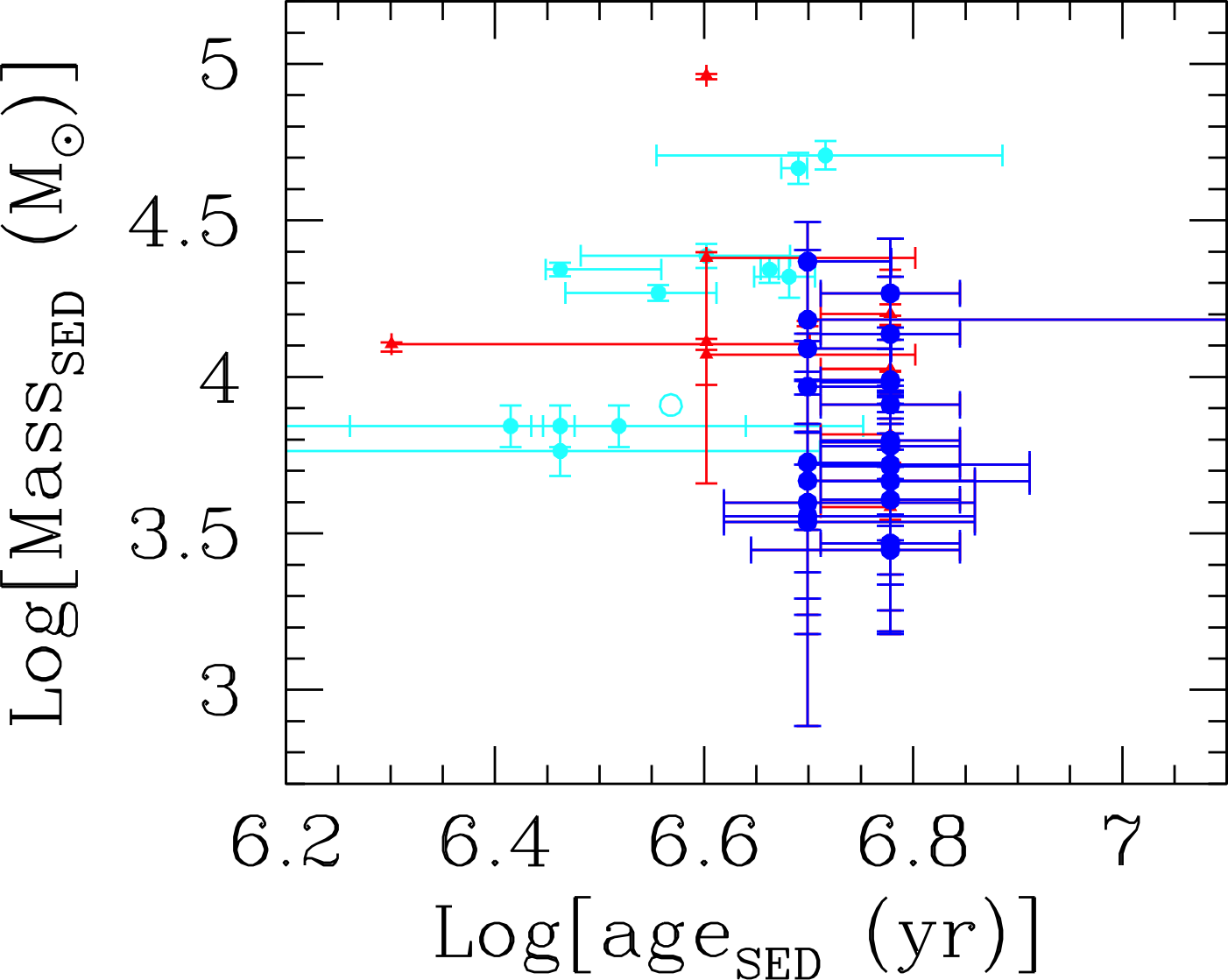}{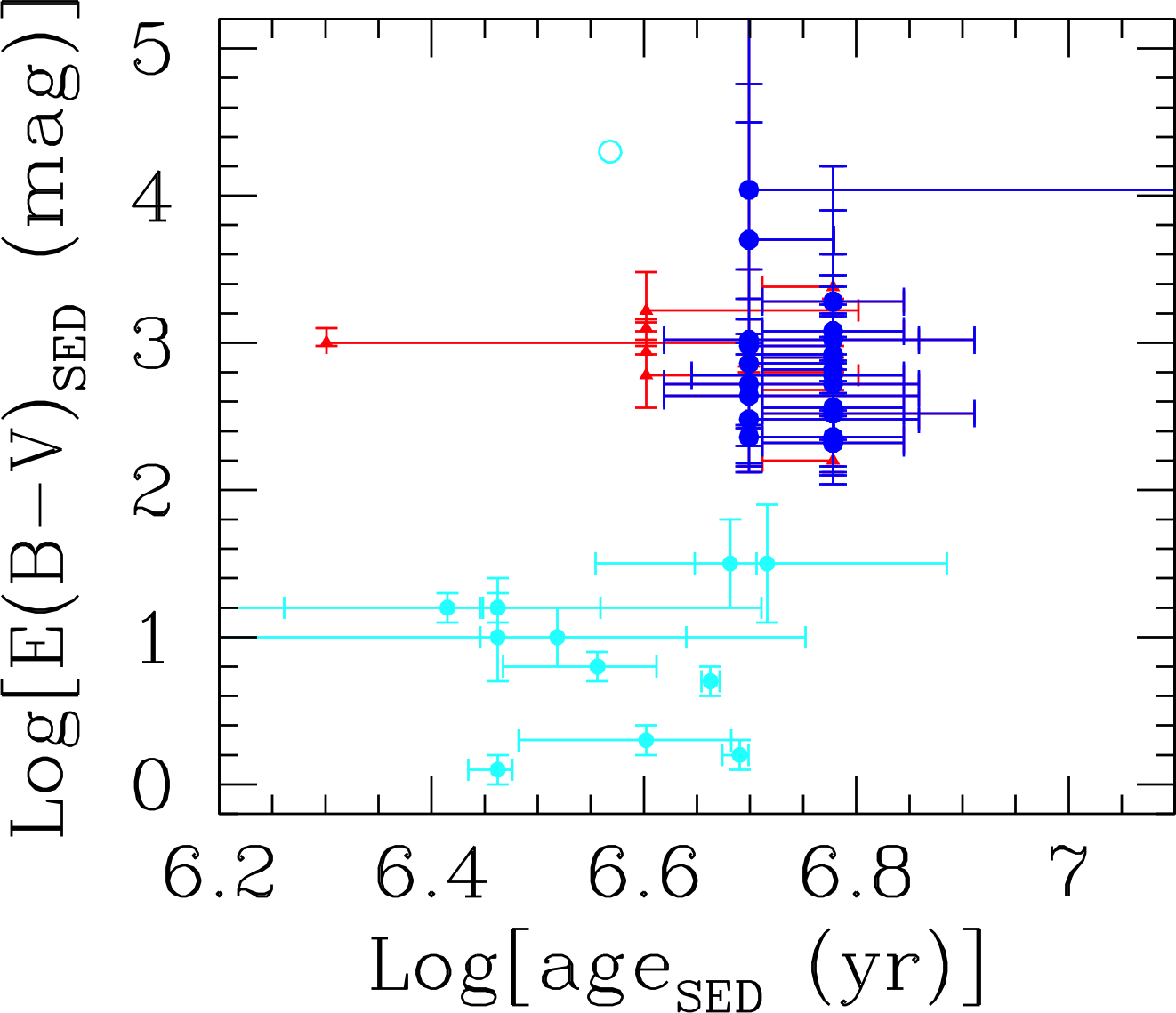}
\caption{(Left:) The stellar mass as a function of the age derived from the SED fits, with 1~$\sigma$ uncertainties. Blue points indicate sources with $\chi^2_{red}\le6$ from the SED fits (see list in Table~\ref{tab:source_properties}), while red points are for sources with larger values of $\chi^2_{red}$. A total of 23 sources, 2/3 of the final sample, have good SED fits according to this criterion. The cyan points show the location, with uncertainties, on this plot of the radio sources from \citet{Reines+2008}, after rescaling to our preferred distance for 
NGC\,4449; the empty cyan circle marks the upper limit in the \citet{Reines+2008} sample. (Right:) The color excess, E(B--V), as a function of age derived from the SED fits. Symbols 
and error bars are as in the left--hand--side panel.}
\label{fig:age_mass}
\end{figure}

\begin{figure}
\plottwo{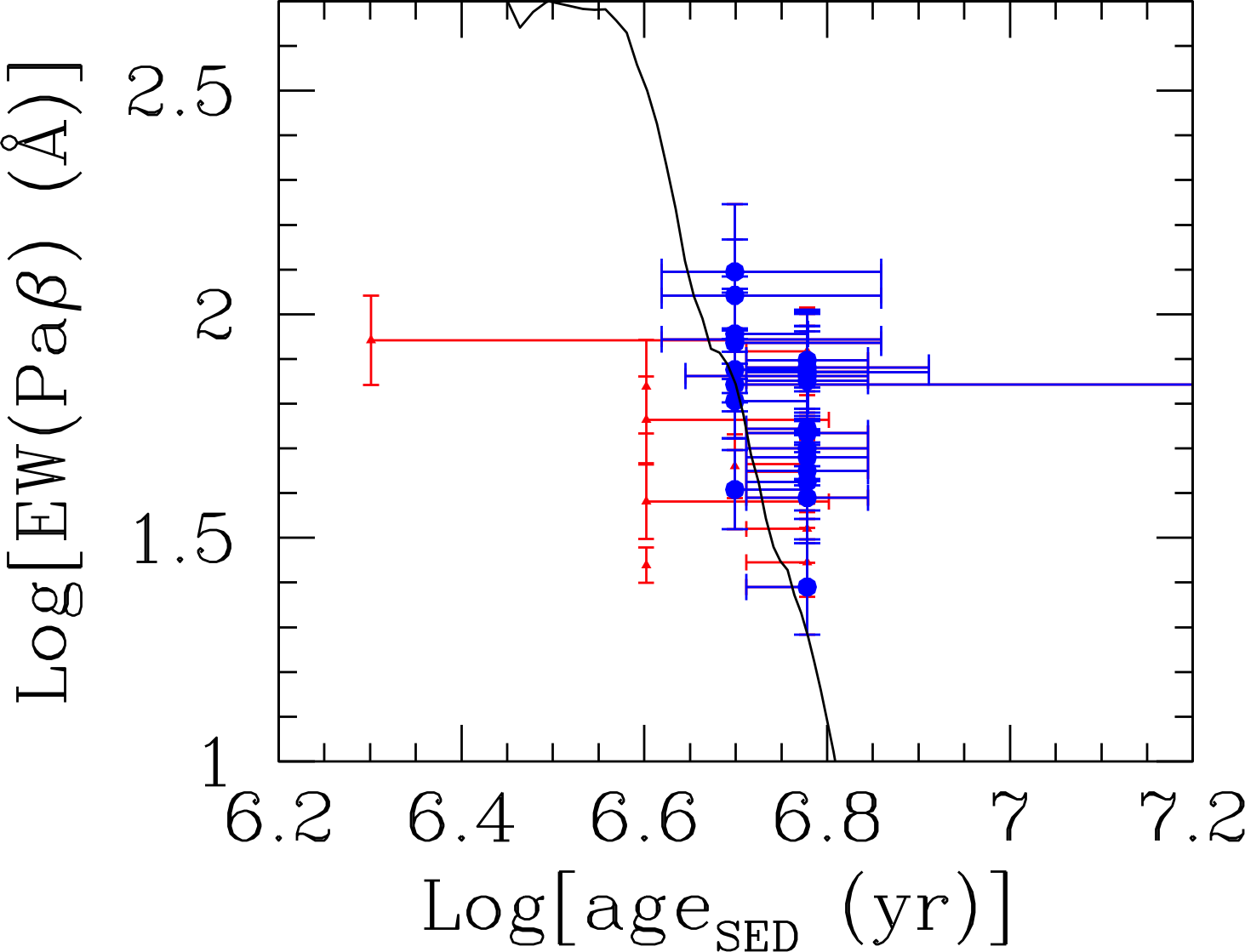}{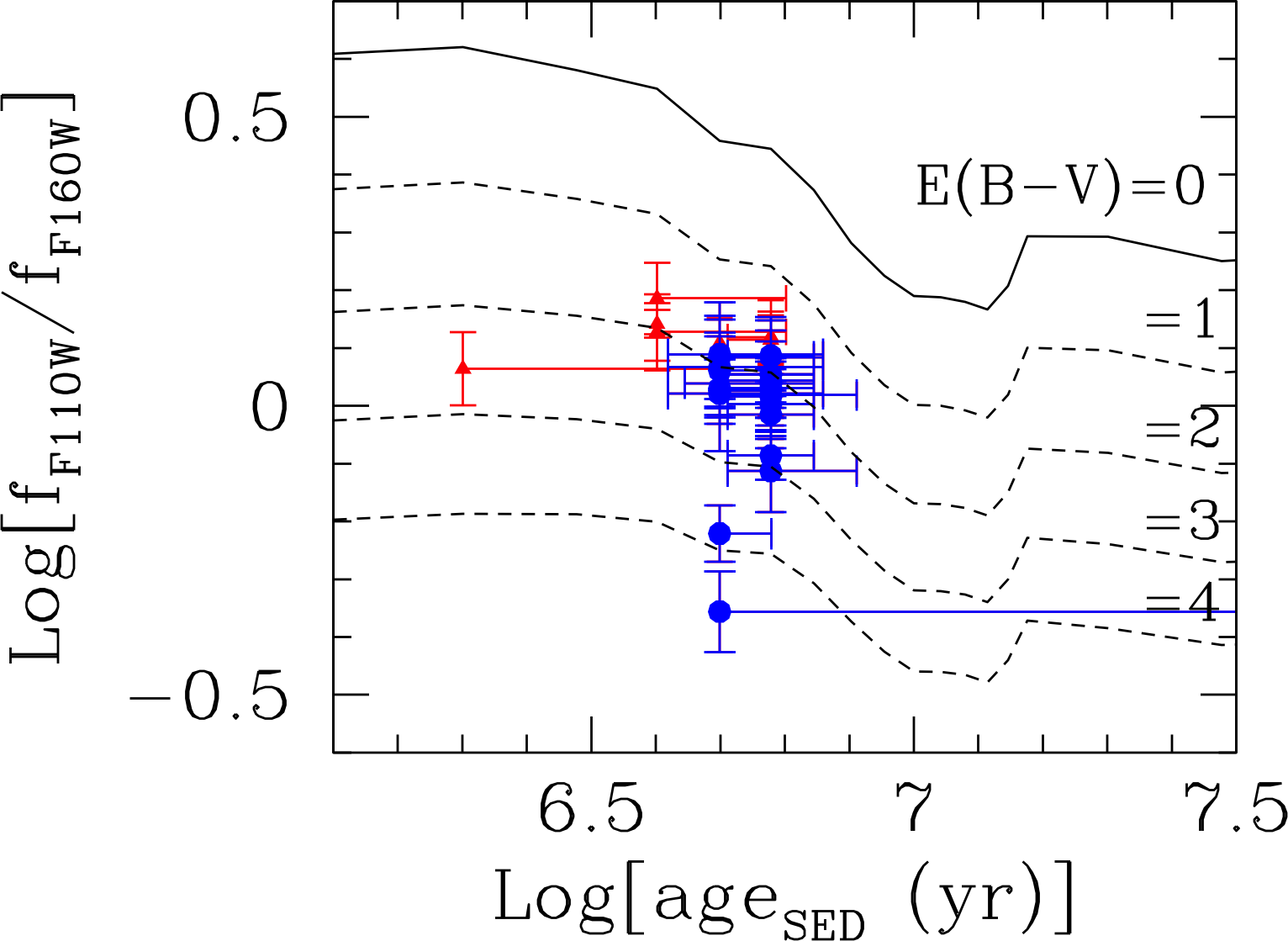}
\caption{(Left:) The EW(Pa$\beta$) as a function of the age derived from the SED fits, with 1~$\sigma$ uncertainties. The data are compared with the models from Starburst99 for Geneva tracks, showing that the SED ages are consistent with the measured line EW.  (Right:) The observed broad--band IR colors J$-$H as a function of the age from the SED fit; symbols are as in the previous panel. Yggdrasil models with Geneva tracks for the colors are shown for increasing values of the color excess E(B$-$V), showing that the data are consistent with large values of the color excess, in agreement with results from the fits. It should be noted that this comparison is somewhat circular, since the colors are used to derive the ages and color excesses. Symbols are as in Figure~\ref{fig:age_mass}.} 
\label{fig:color_age}
\end{figure}

The combination of Geneva models with the LMC extinction curve yields the lowest $\chi^2_{red}$ overall across the 34 regions. Examples of the best--fit SEDs are shown in Figures~\ref{fig:SEDs} and \ref{fig:SEDs_2} for a selection of $\chi^2_{red}$ values. The best--fit age, mass, and color excess for each source, with their $\chi^2_{red}$ values, are listed in Table~\ref{tab:source_properties}. We define as `acceptable' a fit with $\chi^2_{red}\le$6; this is an arbitrary value chosen on the basis of visual inspection of the SED fits, resulting in 23 sources with acceptable fits. The choice of $\chi^2_{red}\le$6 as cutoff has modest impact on our conclusions; choosing $\chi^2_{red}\le$4 would still yield 22 sources with acceptable fits (Table~\ref{tab:source_properties}). The remaining sources range in $\chi^2_{red}$ value from bad to catastrophically bad; these sources will be indicated as `low--significance', with different symbols from the acceptable ones in all Figures that follow. The case of the worst $\chi^2_{red}$ is shown in the right hand--side panel of Figure~\ref{fig:SEDs_2}. The best--fit ages are concentrated around 5--6~Myr, while the masses span the $\sim$3,000--25,000~M$_{\odot}$ range, with a median value of 5,600~M$_{\odot}$ (Figure~\ref{fig:age_mass}, left). These mass values support our use of deterministic models for deriving the physical parameters of the sources. The color excess E(B$-$V) is systematically high, $\gtrsim$2~mag, as expected from earlier considerations based on H$\alpha$ and Pa$\beta$ selection criteria (Figure~\ref{fig:age_mass}, right). The measured EW(Pa$\beta$) as well as the J$-$H colors agree with the model expectations; in particular, the observed J$-$H colors are consistent with stellar populations attenuated by large values of E(B$-$V) (Figure~\ref{fig:color_age}). 

\begin{figure}
\plottwo{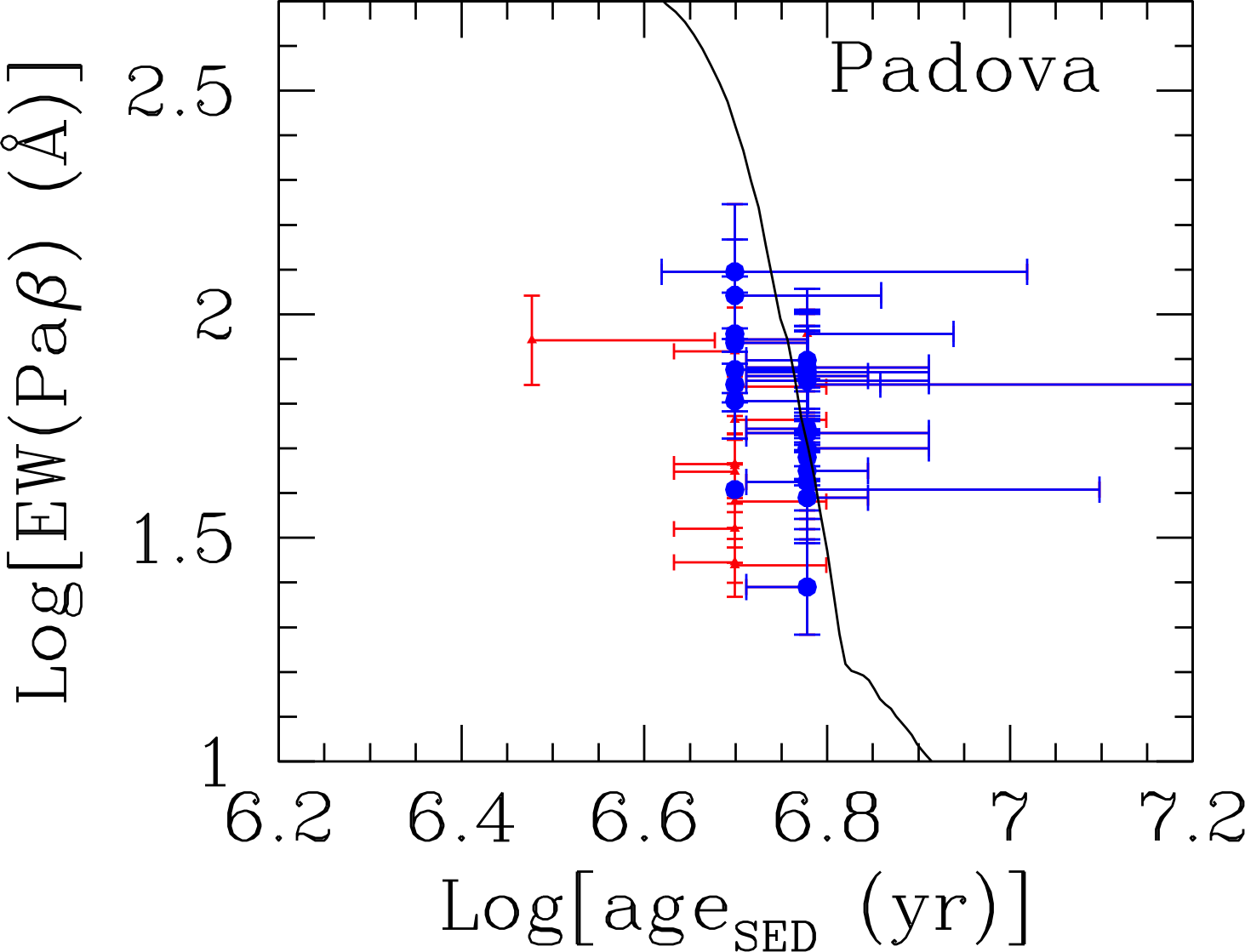}{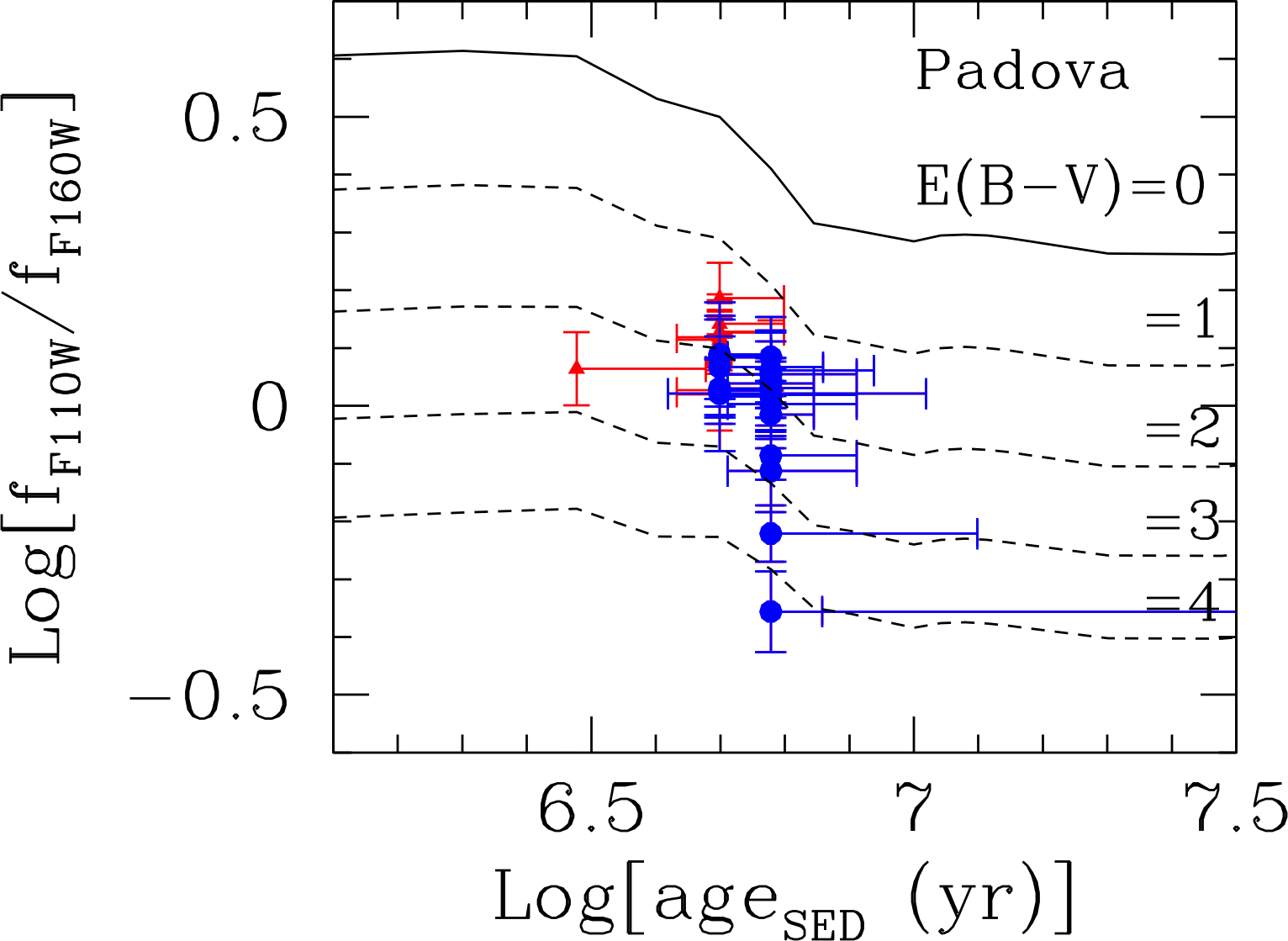}
\caption{The same as Figure~\ref{fig:color_age}, but using the Padova$+$AGB tracks instead of the Geneva tracks. The physical parameters derived for the sources are basically unchanged. Symbols are as in Figure~\ref{fig:age_mass}.} 
\label{fig:color_age_padova}
\end{figure}

Using the  Padova$+$AGB tracks instead of the Geneva ones yields slightly worse $\chi^2_{red}$ values, between a few \% to a factor $\sim$1.5, with a median fraction $\sim$20\% worse. However, the general picture does not change: ages are still around 5--6~Myr, extinctions are high, and masses are large, between a few percent and a factor $\sim$2  larger than for the Geneva tracks. Figure~\ref{fig:color_age_padova} illustrates some of these results.

\begin{figure}
\plottwo{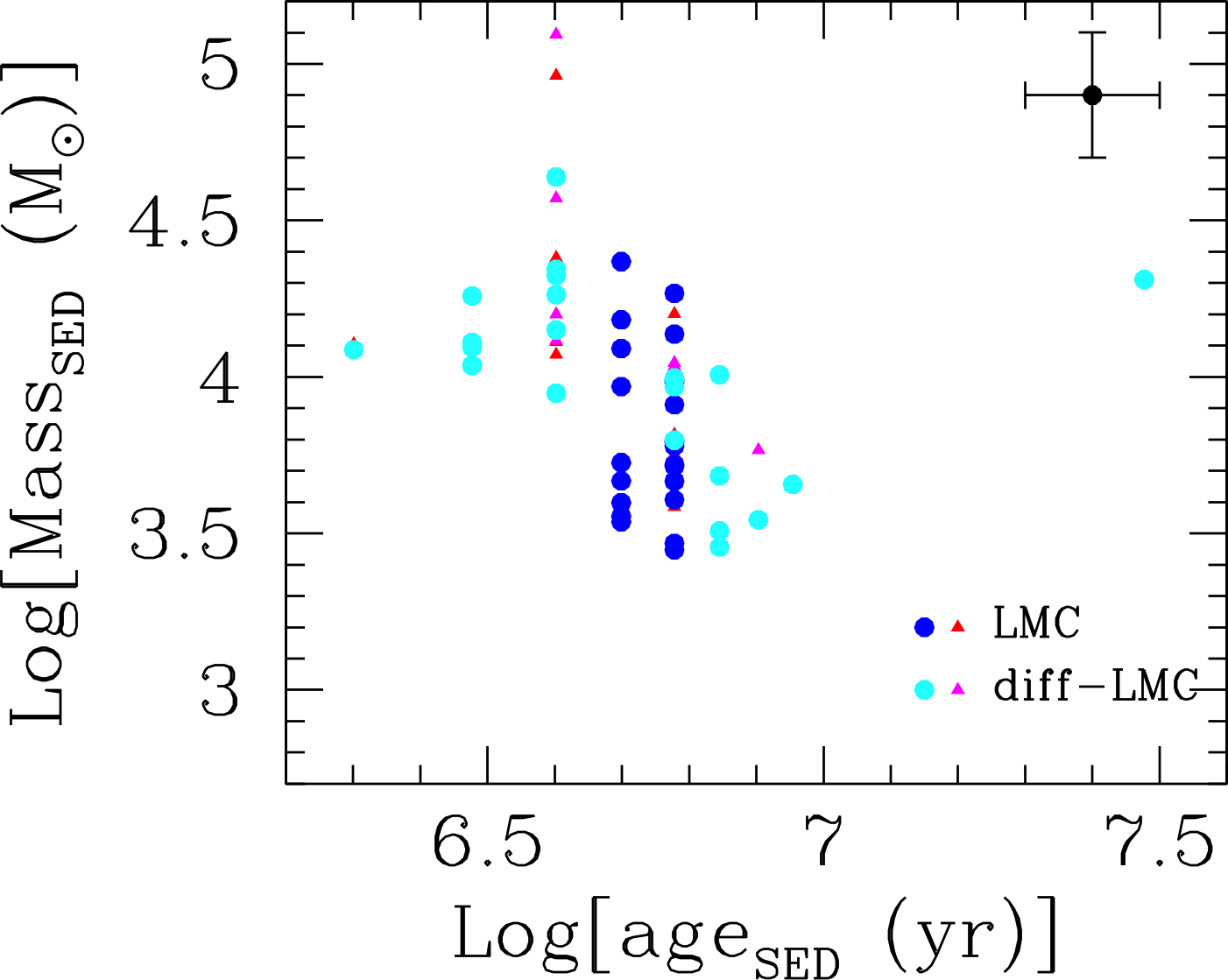}{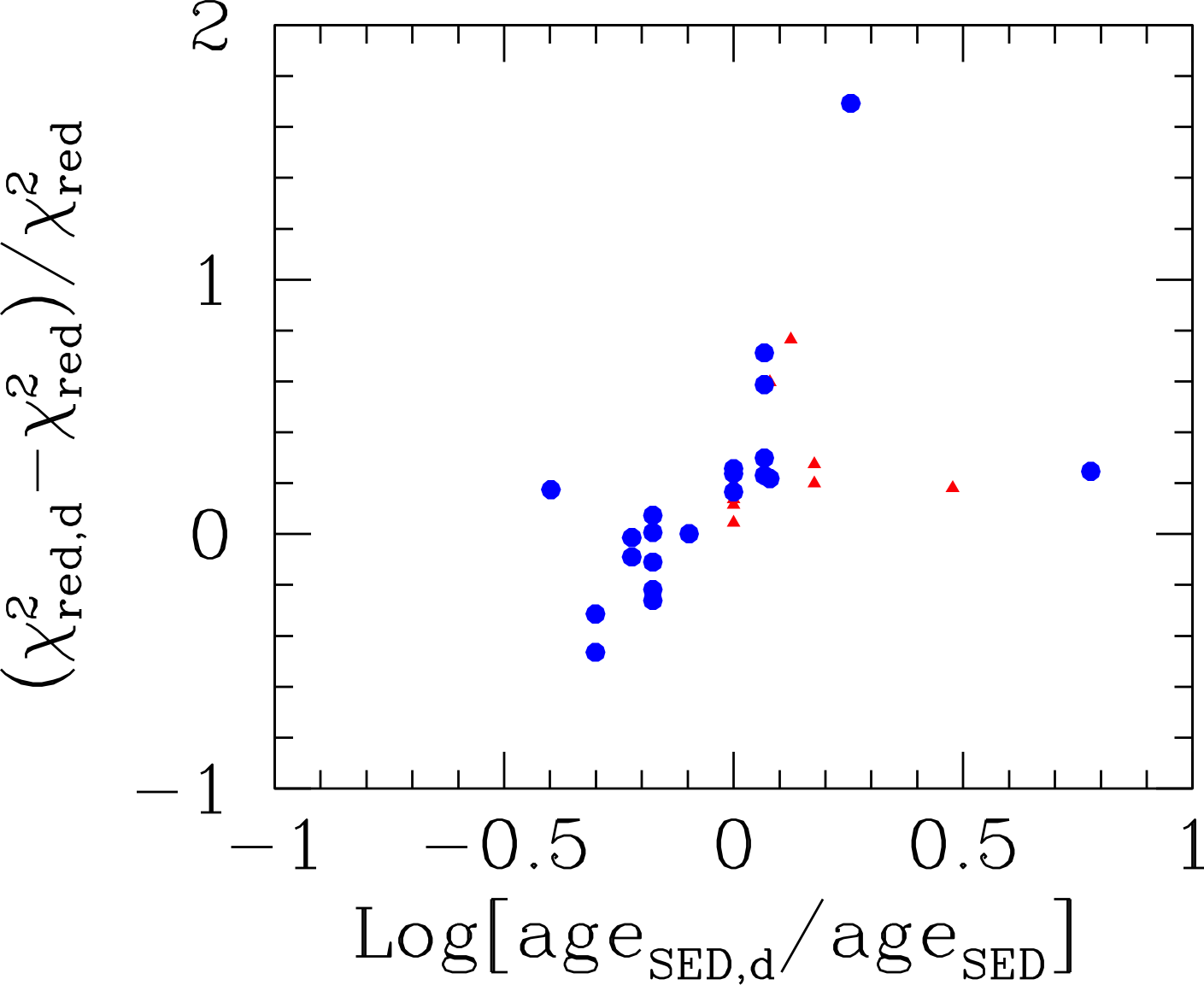}
\caption{(Left:) The best--fit ages and masses for the 34 sample sources derived using the Geneva tracks with the LMC extinction curve (our default model; blue and red symbols for high-- and low--significance fits, respectively) and with the differential LMC extinction for emission lines and stellar continuum (see section~\ref{sec:models}, cyan and magenta symbols for high-- and low--significance fits, respectively). A representative error bar is shown, to aid clarity in the plot. (Right:) The fractional change in the goodness--of--fit as a function of the difference in age between the differential LMC and standard LMC extinctions. The subscript `d' indicates values obtained with the differential LMC extinction curve. Blue and red symbols are used as in previous Figures: blue circles for $\chi^2_{red}\le6$ and red triangles for lower significance.} 
\label{fig:age_mass_diffLMC}
\end{figure}

The combination of Geneva models with the other two dust  attenuation prescriptions considered in this work, i.e., differential--LMC and starburst, yields slightly worse and significantly worse $\chi^2_{red}$ values for the fits, respectively. The $\chi^2_{red}$ resulting from using the starburst curve is at least a factor of 2--3 worse than using the LMC curve, and we drop this option from further consideration. We discuss the differential--LMC case in greater detail, as the $\chi^2_{red}$  discrepancies with the LMC case are relatively small. The masses and ages resulting from the differential LMC case are displayed in Figure~\ref{fig:age_mass_diffLMC} in comparison to our default case of the LMC extinction, to highlight similarities 
and differences. The clearest difference between the two attenuation approaches is that the fits  with the differential LMC extinction yield a wider range of ages for the sources, $\sim$3--8~Myr. In particular we now find 11 sources, i.e., half of the high--significance ones, with ages$\le$4~Myr, as opposed to none with the standard 
LMC extinction. However, in most cases, the values of the $\chi^2_{red}$ for the differential LMC extinction are worse than those for the standard LMC extinction  (Figure~\ref{fig:age_mass_diffLMC}, right); only for 6 sources (3 with age 3~Myr and 3 with age 4~Myr) the goodness--of--fit improves in the case of  the differential LMC extinction. Furthermore, the best--fit E(B$-$V) values are all larger than 2.5~mag for the differential LMC extinction, thus they do not change the overall scenario that these are extremely dust--reddened sources.

One important limitation of our models is the absence of pre-main sequence stars, which are expected to be present at these young ages and contribute to the near--IR continuum emission. \citet{Greissl+2010} explicitly consider the contribution of pre--main sequence stars in their modeling of the near--IR data of a young star cluster in the Antennae galaxy. They conclude that the contribution from these unevolved stars is already small at 3~Myr, and rapidly decreases for increasing age. We thus conclude that the contribution of pre--main sequence stars to our results is negligible. 

For completeness, we also attempt SED fits using synthetic photometry of Red Supergiants (RSGs). RSGs' luminosities in the Large Magellanic Cloud, which has roughly the metallicity of NGC\,4449, can be as bright as 10$^{34}$~erg~s$^{-1}$~\AA$^{-1}$ in the J and H bands, with Log[f(J)/f(H)]$\sim$0.1--0.12 \citep{Oestreicher+1997, Davies+2013}. Both luminosities and colors are in the range of what we observe for our sources (Table~\ref{tab:photometry} and Figure~\ref{fig:color_age}), although RSGs do  not have line emission at 1.28~$\mu$m. We use the model spectra of \citet{Lancon+2007} with a fiducial temperature of 3,500~K, which matches the typical temperatures of LMC RSGs \citep{Davies+2013}. The model spectra are provided at three gravity values \citep{Lancon+2007}, and we use all three to maximize our probability of matching the observed SEDs. We find that the SED fits using these models yield $>$50\% worse goodness--of--fit than the fits using the SSP models, without solving the `too much extinction' problem: the typical source, if they are RSGs, needs to be behind a dust screen with E(B--V)$>$1~mag in order to match our observations. 

In summary, the Geneva tracks with the LMC extinction curve yield the best results in terms of goodness--of--fits as measured from the $\chi^2_{red}$: the sources have ages $\sim$5--6~Myr, masses with median value 5,600~M$_{\odot}$, and foreground extinctions E(B--V)$>$2~mag. About half--dozen sources are better fit with the differential LMC extinction, and have ages that are $\sim$3--4~Myr, masses $>$10$^4$~M$_{\odot}$ and extinctions E(B--V)$>$2.6~mag. As these sources represent only 25\% of our high--significance sample, we will only use the results from the standard LMC extinction fit, listed in Table~\ref{tab:source_properties}, as our fiducial results for the reminder of the paper.

\begin{deluxetable*}{rrrrr}
\tablecaption{Physical Properties of the Sources\label{tab:source_properties}}
\tablewidth{0pt}
\tablehead{
\colhead{ID$^1$} & \colhead{Age$_{SED}$} & \colhead{E(B--V)$_{SED}$} & \colhead{Log(Mass$_{SED}$)} &  \colhead{$\chi^2_{red}$} \\
\colhead{} & \colhead{Myr} & \colhead{mag} & \colhead{M$_{\odot}$} & \colhead{} \\
}
\startdata
  1 & 5.$_{-0.}^{+1.}$ &2.98$_{-0.06}^{+0.02}$ &4.09$_{-0.15}^{+0.02}$ &   3.41\\
  2 & 4.$_{-0.}^{+0.}$ &3.10$_{-0.02}^{+0.04}$ &4.11$_{-0.03}^{+0.01}$ &  14.00\\
  3 & 6.$_{-1.}^{+0.}$ &3.38$_{-0.00}^{+0.08}$ &4.20$_{-0.01}^{+0.14}$ &  11.39\\
  4 & 6.$_{-1.}^{+2.}$ &3.02$_{-0.36}^{+0.52}$ &3.72$_{-0.38}^{+0.27}$ &   0.39\\
  5 & 4.$_{-0.}^{+2.}$ &3.22$_{-0.20}^{+0.06}$ &4.38$_{-0.41}^{+0.02}$ &  15.48\\
  6 & 6.$_{-1.}^{+0.}$ &2.82$_{-0.00}^{+0.10}$ &3.98$_{-0.03}^{+0.14}$ &   3.26\\
  7 & 6.$_{-1.}^{+1.}$ &2.36$_{-0.20}^{+0.16}$ &3.80$_{-0.24}^{+0.14}$ &   1.47\\
  8 & 5.$_{-0.}^{+1.}$ &2.36$_{-0.24}^{+0.32}$ &3.73$_{-0.21}^{+0.10}$ &   0.37\\
  9 & 6.$_{-1.}^{+1.}$ &2.72$_{-0.22}^{+0.44}$ &3.78$_{-0.26}^{+0.19}$ &   0.97\\
 10 & 6.$_{-1.}^{+0.}$ &2.92$_{-0.04}^{+0.22}$ &3.99$_{-0.04}^{+0.17}$ &   2.69\\
 11 & 5.$_{-0.}^{+0.}$ &2.80$_{-0.00}^{+0.04}$ &4.17$_{-0.01}^{+0.01}$ &  10.07\\
 12 & 6.$_{-1.}^{+0.}$ &2.68$_{-0.00}^{+0.10}$ &4.03$_{-0.01}^{+0.14}$ &  10.21\\
 13 & 5.$_{-0.}^{+1000}$ &4.04$_{-1.74}^{+0.62}$ &4.18$_{-0.81}^{+0.22}$ &  1.12\\
 14 & 6.$_{-1.}^{+1.}$ &2.52$_{-0.40}^{+0.28}$ &4.14$_{-0.27}^{+0.18}$ &   3.45\\
 15 & 4.$_{-0.}^{+0.}$ &2.94$_{-0.02}^{+0.02}$ &4.96$_{-0.01}^{+0.00}$ &  91.32\\
 16 & 6.$_{-1.}^{+1.}$ &3.28$_{-0.62}^{+0.30}$ &4.27$_{-0.35}^{+0.18}$ &   2.72\\
 17 & 4.$_{-0.}^{+2.}$ &2.78$_{-0.22}^{+0.14}$ &4.07$_{-0.41}^{+0.03}$ &  10.19\\
 18 & 6.$_{-1.}^{+1.}$ &2.32$_{-0.22}^{+0.20}$ &3.71$_{-0.24}^{+0.17}$ &   1.71\\
 19 & 6.$_{-1.}^{+0.}$ &2.20$_{-0.04}^{+0.08}$ &3.58$_{-0.04}^{+0.14}$ &   8.02\\
 20 & 6.$_{-1.}^{+2.}$ &2.52$_{-0.48}^{+0.46}$ &3.67$_{-0.41}^{+0.25}$ &   0.64\\
 21 & 6.$_{-1.}^{+0.}$ &3.02$_{-0.04}^{+0.24}$ &3.82$_{-0.04}^{+0.42}$ &  10.76\\
 22 & 6.$_{-1.}^{+1.}$ &2.56$_{-0.22}^{+0.40}$ &3.61$_{-0.24}^{+0.21}$ &   2.66\\
 23 & 2.$_{-0.}^{+2.}$ &3.00$_{-0.02}^{+0.08}$ &4.10$_{-0.02}^{+0.00}$ &  60.38\\
 24 & 5.$_{-1.}^{+2.}$ &2.72$_{-0.28}^{+0.30}$ &3.54$_{-0.36}^{+0.31}$ &   2.69\\
 25 & 6.$_{-1.}^{+0.}$ &2.80$_{-0.00}^{+0.10}$ &4.02$_{-0.01}^{+0.14}$ &  16.19\\
 26 & 5.$_{-0.}^{+2.}$ &2.48$_{-0.30}^{+0.22}$ &3.67$_{-0.38}^{+0.05}$ &   1.15\\
 27 & 6.$_{-1.}^{+0.}$ &2.90$_{-0.02}^{+0.12}$ &3.79$_{-0.02}^{+0.15}$ &   4.55\\
 28 & 5.$_{-1.}^{+2.}$ &3.02$_{-0.86}^{+0.88}$ &3.55$_{-0.67}^{+0.43}$ &   1.01\\
 29 & 6.$_{-1.}^{+1.}$ &3.08$_{-0.22}^{+0.30}$ &3.91$_{-0.24}^{+0.18}$ &   2.43\\
 30 & 5.$_{-0.}^{+1.}$ &3.70$_{-0.40}^{+0.40}$ &4.37$_{-0.23}^{+0.13}$ &   1.14\\
 31 & 5.$_{-0.}^{+1.}$ &2.86$_{-0.06}^{+0.22}$ &3.97$_{-0.05}^{+0.14}$ &   1.51\\
 32 & 5.$_{-1.}^{+2.}$ &2.64$_{-0.22}^{+0.64}$ &3.60$_{-0.36}^{+0.39}$ &   1.20\\
 33 & 6.$_{-1.}^{+1.}$ &2.32$_{-0.28}^{+0.44}$ &3.47$_{-0.29}^{+0.24}$ &   0.52\\
 34 & 6.$_{-2.}^{+1.}$ &2.78$_{-0.24}^{+0.24}$ &3.45$_{-0.26}^{+0.40}$ &   1.89\\
\enddata
\tablecomments{Age (Myr), color excess E(B--V) (mag), logarithm of the stellar mass (M$_{\odot}$), and reduced $\chi^2$ from the SED fit of the photometry in Table~\ref{tab:photometry} for  the 34 sources in our sample.}
\end{deluxetable*}

\section{Discussion} \label{sec:discussion}

\subsection{What are these sources?}

The ages  and masses from SED fits and the presence of hydrogen recombination line emission make our sources consistent with being young star clusters, younger than $\sim$7--8~Myr. The requirement that the Pa$\beta$ emission be detected with S/N$\gtrsim$3 places a constraint on its EW, and thus on the maximum age of the sources: the minimum EW$\sim$25~\AA\  translates into a maximum age of about 6--7~Myr \citep{Leitherer+1999}. For reference, beyond 7~Myr the ionized gas emission is between $>$30 
(Padova$+$AGB) and $>$50 (Geneva) times fainter than at 3~Myr.  

The existing studies on the  star cluster population of  NGC\,4449 center on optically--detectable sources \citep{Whitmore+2020}, and are thus complementary  to our selection of optically--faint sources. We compare our sources with the optically--bright clusters in a later subsection. \citet{Sokal+2015} identify a partially--embedded, Wolf--Rayet star cluster in the galaxy; the presence of Wolf--Rayet features gives this cluster a minimum age of 3~Myr and indicate that stellar winds are active in NGC\,4449. This cluster would provide an important comparison for our sources, but it falls outside the FoV of our WFC3/IR images. The closest study to ours is the one by \citet{Reines+2008}, where the authors focus on radio--selected sources.

\begin{figure}
\plotone{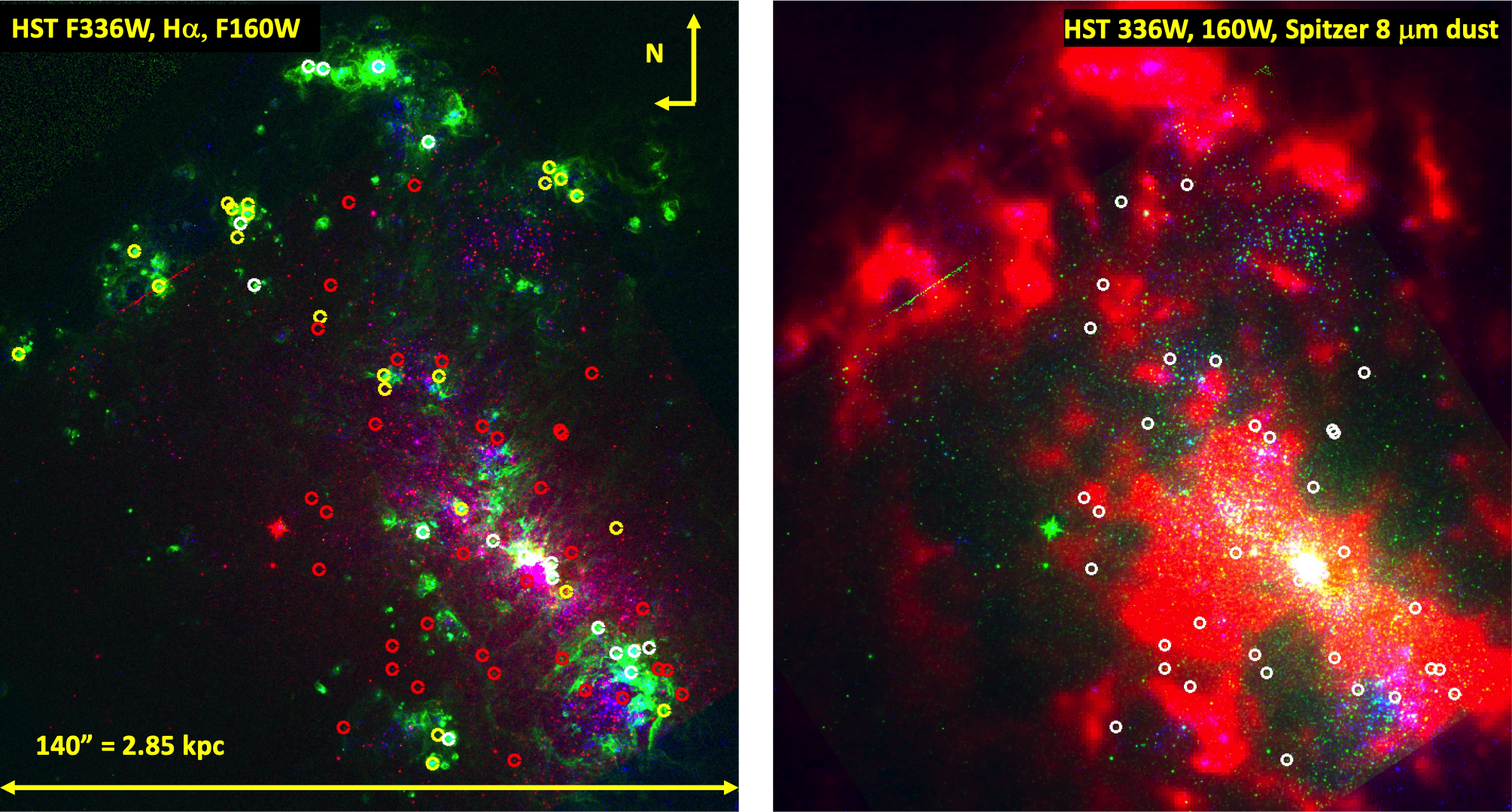}
\caption{(Left:) A three--color composite of NGC\,4449 using F336W (blue),  continuum--subtracted H$\alpha$+[NII] (green), and F160W (red). The FoV is 2.5$^{\prime}\times$2.5$^{\prime}$ (3.05$\times$3.05~kpc$^2$), slightly larger than the FoV of WFC3/IR to show the location of all of the \citet[][]{Reines+2008}'s sources. The 39 \citet[][]{Reines+2008}'s sources were identified from multi--wavelength VLA data (see text) and include thermal/likely thermal/mixed sources (white circles) and non--thermal/uncertain sources (yellow circles). Our sources, identified from the stellar--continuum subtracted Pa$\beta$ WFC3/IR image, are shown as red circles. All circles have a radius of 1$^{\prime\prime}$.0 to facilitate visualization. North is up, East is left. (Right:) A three--color composite of HST (F336W=blue and F160W=green) and Spitzer 8~$\mu$m dust (red) images, showing the location of our sources (white circles) relative to the location of the dust emission. This figure shows a larger dynamical range for the 8~$\mu$m dust emission than Figure~\ref{fig:galaxy}, to highlight that our sources are generally located at  the margins of emission peaks.}
 \label{fig:radio_sources}
\end{figure}

\citet{Reines+2008} identified  39 compact sources in their VLA observations at 1.3 cm, 3.6 cm, and 6 cm of  NGC\,4449. 
Of these 39 sources, 13 were classified as thermal or `likely thermal', 4 are mixed,  and the remaining ones are either non--thermal or uncertain. 
The 39 sources are shown in Figure~\ref{fig:radio_sources}, left panel, with white circles for the thermal/likely thermal/mixed sources and yellow circles for the other sources. Our sources are located in correspondence of the red circles on the same figure. The radio sources are clearly separated from the dust--buried sources in our sample. They 
are generally located in areas of intense H$\alpha$ emission, while the dust--buried sources are,  by design, located at the edges of or away from the brightest line emission regions.

\citet{Reines+2008} model the 13 thermal and likely thermal sources in their sample using multi--wavelength data from the ultraviolet to the radio. The 
sources have ages in the range $\sim$2.5-5.5~Myr with a median of 3.6~Myr  and masses in the range $\sim$(6--51)$\times$10$^3$~M$_{\odot}$, with a median of 
16$\times$10$^3$~M$_{\odot}$, after rescaling to our preferred distance (cyan datapoints in Figure~\ref{fig:age_mass}). Their sources on average are a factor $\sim$1.5 younger and 
$\sim$3 more massive than ours. These authors observe a mass--age  anticorrelation, typical of luminosity--limited observations subject to 
size--of--sample effects \citep{Hunter+2003}. Separating the radio thermal sources in two groups: younger and older than 4~Myr, the younger group 
includes 8 sources with median age 2.9~Myr and median mass 7$\times$10$^3$~M$_{\odot}$, while the older group is comprised of 5 sources 
with median age 4.8~Myr and median mass 24$\times$10$^3$~M$_{\odot}$. The older group is closer in age to our dust--buried sample, and is a factor 
of $\gtrsim$4 more massive (Figure~\ref{fig:age_mass}, left panel). That our sources are less massive, on average, than the \citet{Reines+2008} sources when considering equal--age bins should not be surprising. This reflects, again, a combination of size--of--sample effect \citep{Hunter+2003} and sensitivity limits of the radio data. In addition, the selection function we apply to our sources may potentially remove the most massive clusters (see next subsection).

\begin{figure}
\plotone{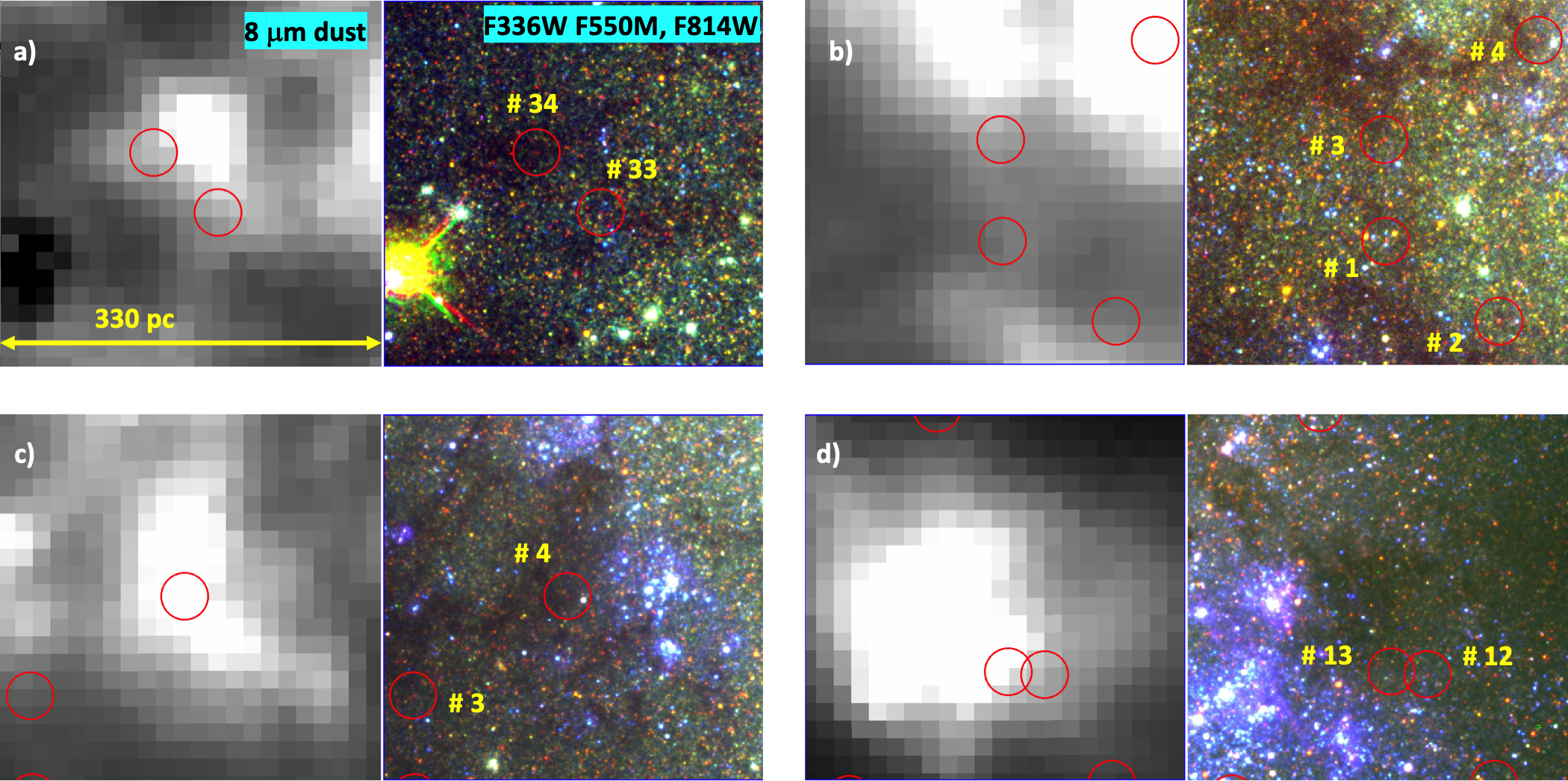}
\caption{The location of several of our sources is shown in four separate panels (a, b, c, d) displaying cut--outs of the 8~$\mu$m dust map (left of each  panel, in b/w) and of a three--color composite using F336W (blue), F550M (green), and F160W (red, right of each panel). All cut--outs have size of  330 pc. The three HST continuum bands are chosen to explicitly avoid any line emission, to highlight the location of dust lanes and clouds. Our sources are shown as red circles, marked with the ID number from Table~\ref{tab:photometry}. Source \# 4 is shown twice, both in panels b and c, to better highlight the dusty region it is sitting on. All circles have a radius of 1$^{\prime\prime}$.0 to facilitate visualization. North is up, East is left.}
 \label{fig:individual_cutouts}
\end{figure}

Several  of the sources in our sample are located a few hundred pc away from the closest areas of bright H$\alpha$ and Pa$\beta$ emission; they are unlikely to have 
formed in these more (optically) active regions, since they would not have had the time to travel that far from their natal site. Assuming a (generous) speed of 10~km/s, the typical 
source would have traveled about 50--60~pc in 5--6~Myr, less than their distance from the more optically  active regions. Furthermore most sources are located at the 
periphery of areas where the optical light is depressed but the dust emission is strong, marking the location of dust clouds (Figures~\ref{fig:radio_sources}, right, and \ref{fig:individual_cutouts}). Our star clusters are, thus, likely to be in the process of emerging from these dust clouds. They may be the `tip of the iceberg'  of larger numbers of clusters that are forming within the IR--bright dust clouds and are undetected because of the large dust column densities. These completely hidden sources will be primary targets for JWST studies\footnote{JWST Cycle~1 imaging for NGC\,4449 is part of program \# 1783.}. 

In summary, our sources are likely to be relatively young star clusters born in--situ and still affected  by significant dust attenuation, A$_V>$6~mag. These clusters are young, 
but not extremely young, clustering around an age of 5--6~Myr. 

\subsection{Where are the youngest sources?}

Our selection criteria place strong constraints on the type of sources we are likely to isolate. The constraint on the colors excess, E(B$-$V)$\gtrsim$2.1~mag, has been already discussed in section~\ref{sec:selection}. As discussed earlier, selection effects on the EW(Pa$\beta$) limit our maximum age to be younger than 6--7~Myr. 
Below this age limit, we would expect roughly a constant number of clusters in each age bin, if star formation has continued at the same level over the past 6-10~Myr; yet only one cluster candidate, out of 34, has an age consistent with 2~Myr, with a poor fit to its SED (Table~\ref{tab:source_properties}). Even adopting the results from the differential LMC  extinction fits for the cases where this choice yields lower values of $\chi^2_{red}$ then the standard LMC extinction, we find six sources with ages between 3 and 4 Myr, and none with a best--fit age lower than 3~Myr. 

There are several concomitant selection effects that are likely preventing us from securing the youngest, dust--buried sources. Sources younger than 3~Myr in the same mass and color excess range as the sources we select (Table~\ref{tab:source_properties}) would be intrinsically brighter in line emission: a 2~Myr  old star cluster has a luminosity in the hydrogen recombination lines that is about 10 times brighter than a 5~Myr old star cluster, at constant mass \citep{Leitherer+1999}. Thus, a 2~Myr old, 10$^{3.7}$~M$_{\odot}$ cluster would be detectable in H$\alpha$, violating one of our selection criteria, which require marginal or non--detection in the H$\alpha$ line above the local diffuse emission. In order to remain within our selection criteria, the younger cluster would need to be about 10 times less massive than the older one. This implies that 2~Myr old clusters would need to have masses in  the range 300--2500~M$_{\odot}$, with median mass $\sim$500~M$_{\odot}$. Star clusters with these masses are subject to strong stochastic sampling, and highly likely to lack massive, ionizing stars \citep{Fumagalli+2011, Krumholz+2015}. A factor 10 decrease in mass, from $\sim$5,000~M$_{\odot}$ to $\sim$500~M$_{\odot}$ corresponds to a factor $\sim$3 increase in uncertainty in the ionizing photon flux, from $\sim$30\% to $\sim$100\% \citep{Cervino+2002}.  However, a 500~M$_{\odot}$, 2~Myr old cluster would also violate the constraints on the continuum photometry in the J and H bands, since a cluster's intrinsic luminosity does not change significantly in these bands between 2~Myr and 5~Myr; thus, a 500~M$_{\odot}$ cluster would be too faint to fit the observed J and H fluxes. In other words, clusters younger than the ages we derive, $\sim$5--6~Myr, still need to have masses with a median $\sim$5000~M$_{\odot}$ to fit the broad--band photometry. 

Alternatively, a 2~Myr old, 10$^{3.7}$~M$_{\odot}$ star cluster could become undetectable in H$\alpha$ if it is subject to a larger amount of dust extinction than we derive from SED fitting. Depressing the H$\alpha$ luminosity by an additional order of magnitude requires A$_V$ to increase from a median value of 8.6~mag to a median value of $\sim$11.7~mag. In this case, the observed flux in the I, J and H bands would decrease by factors 5.4, 2.5 and 1.8, respectively, lowering the median I--band luminosities to values that have S/N$\lesssim$1. This is too low of a limit for our visual classification to recognize the sources as real detections in that band, hence those sources would not have entered our sample. Young clusters more massive than 10$^{3.7}$~M$_{\odot}$ would proportionally require higher A$_V$ values to become undetectable in H$\alpha$, making them also too faint in the I band. In summary, the combination of our detection limits and selection criteria causes sources younger than 3--4~Myr to be excluded from our sample. Thus, we are not excluding their existence in the galaxy, just their presence in our sample.

An additional selection effect comes from our detection requirements for {\em both} H$\alpha$ and Pa$\beta$, meaning that while we want $H\alpha$ to remain undetected, we still require Pa$\beta$ to be detected with S/N$\ge$3 and to be sufficiently isolated from neighboring regions as to  provide a reliable photometric measurement. This requirement excludes the most crowded (in nebular emission) regions in the galaxy, which are also the brightest and where we would expect the youngest and most massive clusters to be located. Therefore, dust--buried regions younger than 3--4~Myr may be located in areas excluded from our search, further reinforcing the conclusion above that they are likely to be present in the galaxy, but not included in our sample. The comparison with the radio sources detected by \citet{Reines+2008}, discussed  in the previous subsection, lends support to this scenario.  Given all of this, it is perhaps not surprising that we do not isolate very young, dust--buried star clusters.

\begin{figure}
\plottwo{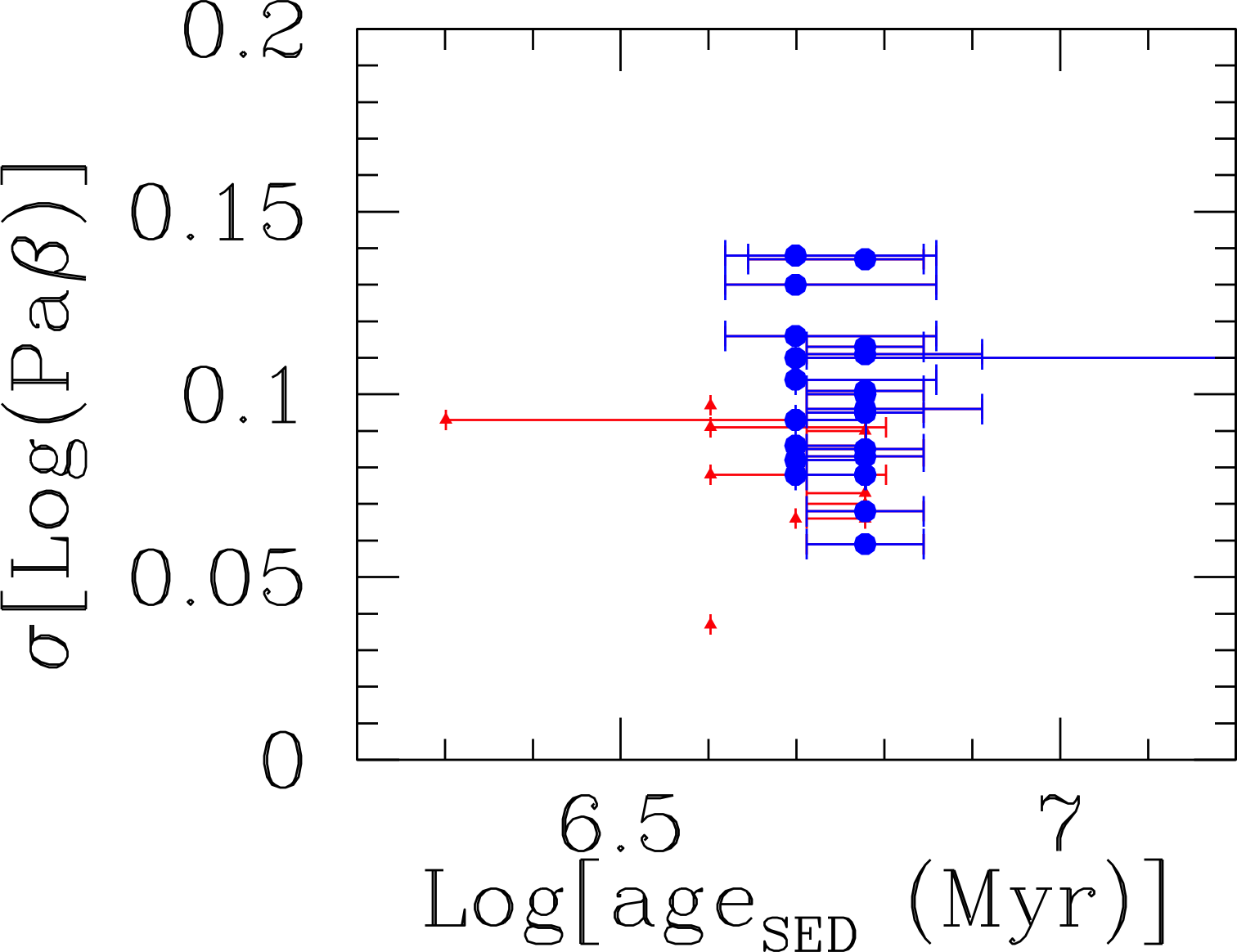}{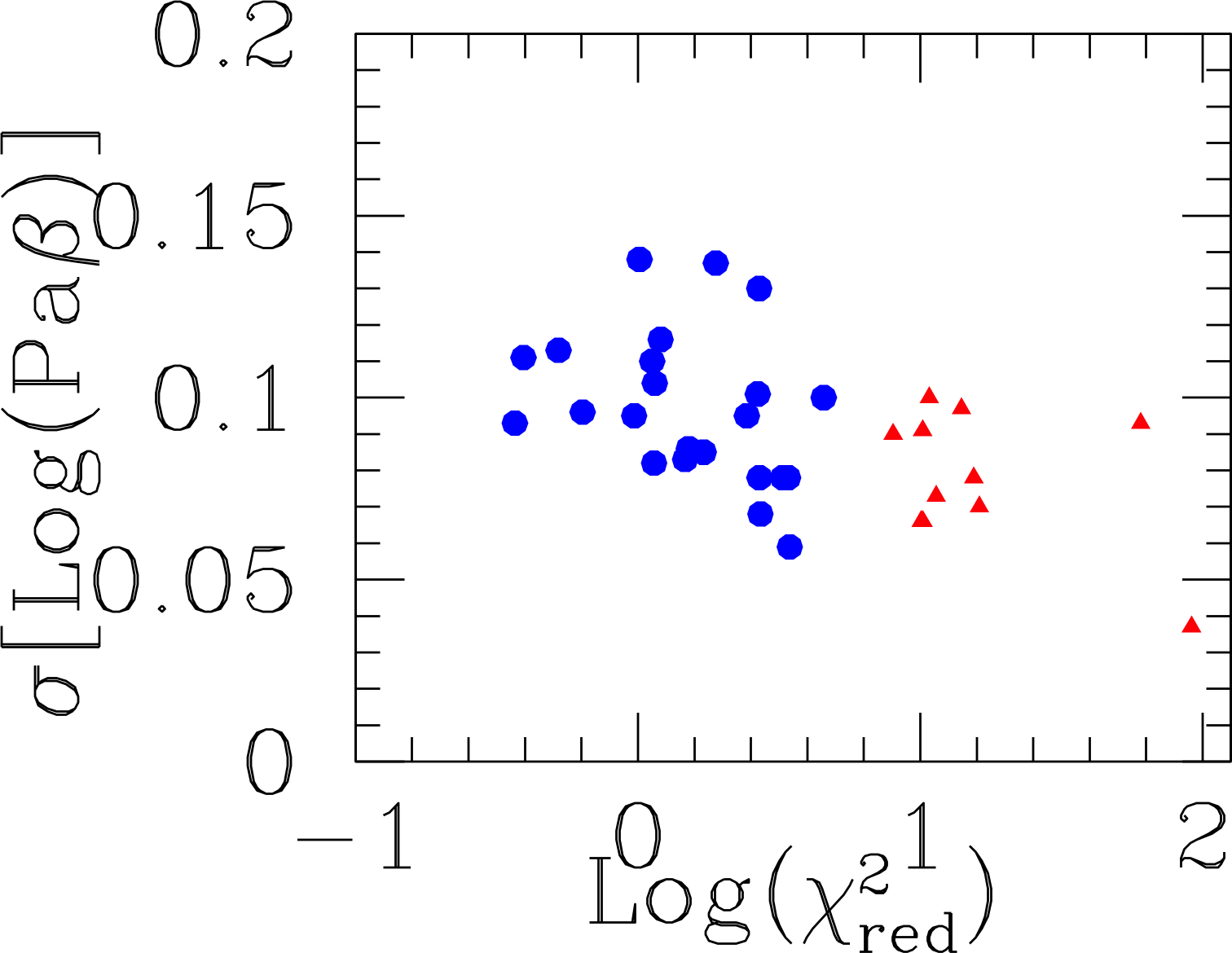}
\caption{The measurement uncertainty in the Pa$\beta$ luminosity as a function of the best--fit age, age$_{SED}$ (with its 1~$\sigma$ uncertainty, left), and of the $\chi^2_{red}$ value of the fits (right), both from Table ~\ref{tab:source_properties}.} 
\label{fig:Pabeta_chi}
\end{figure}

\subsection{Could These Be Spurious Sources or Spurious Ages?}

Because these are unusual sources, the potential that they might be spurious sources needs to be considered. As discussed in section~\ref{sec:selection}, our sources are unlikely to result from artifacts in the stellar continuum subtraction of the near--IR narrow--band filter (F128N). Visual inspection of the continuum--subtracted F658N image enables us to exclude over--subtraction in this image  that would generate artificial `non--detections' in H$\alpha$. 

We also exclude the possibility that these could be background galaxies. The I--band magnitude range of our sources is  I$_{AB}$=20.6--25.6 mag; thus the faint end overlaps with the bright end of the V--band dropouts at redshift z$\sim$5 \citep{Stark+2009}. However, our Pa$\beta$ line  emission would correspond to a line emission at restframe wavelength $\lambda_{rest}\sim$0.21~$\mu$m in a z$\sim$5 galaxy. No known emission line exists at this wavelength, which allows us to exclude high-redshift interlopers. 

The SED fits could  be biased, if, for instance, the uncertainties in the F128N filter, which is also the shallowest among the near--IR ones, were to drive the ages from the fits or the goodness of fit. We test this by plotting the uncertainties on the Pa$\beta$ luminosities as a function of both the ages and the $\chi^2_{red}$ values (Figure~\ref{fig:Pabeta_chi}). We see from the plots that while there is no correlation between ages and uncertainties in Pa$\beta$, $\sigma$[Log(Pa$\beta$)], there is a weak trend between $\chi^2_{red}$  and $\sigma$[Log(Pa$\beta$)], in the sense that larger uncertainties yield better fits. However, this trend is extremely weak and only  driven by a small number of data: the 1~$\sigma$ vertical scatter on the high--significance data ($\gtrsim$0.04 two--sided) is comparable with the dynamical range of the data ($\sim$0.08);  the majority  of the data do not present a significant correlation between the two quantities, implying that smaller uncertainties do not correspond  necessarily to worse SED fits. 

Although we  have excluded the impact of stochastic sampling of the stellar IMF in our analysis, we reconsider this possibility here. The reason is because our SED--fit ages are mainly constrained by the presence of Pa$\beta$ emission, i.e., by the presence of massive, ionizing stars ($>$10--15~M$_{\odot}$). If, for any reason, our  assumption that the IMF is fully sampled is not correct, the ages would be different from those we derive. We consider the case that the sources we select happen to be all sources that have failed to  produce the most massive stars; in other words, we still need at  least a  few stars with mass $>$15~M$_{\odot}$ to  be present in order to generate the hydrogen line emission, but above this mass the IMF is  not fully sampled. 
In this scenario, the hydrogen recombination lines would be intrinsically less bright than those of a fully sampled IMF all the  way to 120~M$_{\odot}$. The consequence is that our   dust--buried sources could be younger, or span a  broader range of ages, than what we derive from the SED fits. The counterargument is that the spatial distribution of the dust--buried sources  is different from that of the low--dust sources discussed in the next subsection, although they span a similar range in both age and mass.  Thus, we would need to invoke a physical mechanism that, at the same time, segregates star clusters spatially within the galaxy and in massive star content. In the absence of an obvious mechanism for this, we discard this possibility, at  least for the moment.

We now discuss our model assumption of a 50\% gas covering factor, i.e., 50\% of ionizing photons absorbed by the gas surrounding the cluster. For our dust--buried sources, we may expect a higher gas covering factor, since an HII region placed within a dense environment can be expected to be compact. If we choose 100\% covering factor (YGGDRASIL only offers 0\%, 50\%, and 100\% covering factors), we find that the best fit ages increase slightly, from a median value of $\sim$5.5~Myr to a median value of $\sim$6~Myr; in addition, all high--significance ($\chi^2_{red}\le 6$) sources now are 6~Myr old. This is easily understood by remembering that our ages are mainly determined by the EW(Pa$\beta$); higher gas covering factors translate into brighter emission lines and larger EWs in the models, implying that older ages are required to match the observations. The average color excess decreases only slightly, by about 0.05~mag, which is understood by recalling that in our fits the color excess value is mainly determined by the shape of the stellar continuum. The derived stellar masses also decrease a little on average, by $\sim$0.15~dex, but non--uniformly across the clusters; the new masses are still in the range 3,000--25,000~M$_{\odot}$. Finally, there is no change in the number of high/low significance fits: the high--significance fits with the 50\% covering factor models remain high--significance with the 100\% covering factor models. Thus, covering factors in the range 50\%--100\% yield almost identical results for the best fit parameters of our dust--buried star clusters. 

We test the impact of our choices of population models and fitting approach by running the stellar population inference code Prospector \citep{Johnson+2021} on the photometry of our sources. While Prospector may not be the best choice for single stellar populations at the extremely young ages of our sources, it represents, nevertheless, a completely different approach, both in terms of fitting and of stellar population libraries \citep{Conroy+2009, ConroyGunn2010, Falcon+2011, Dotter+2016, Choi+2016, Speagle+2020}. We use flat priors in mass, extinction, and age, spanning the full physical range  for clusters, but fix the metallicity at 40\% solar, as appropriate for NGC\,4449. The output physical parameters track well our main results, with the main discrepancies between the  two approaches found for our low--significance clusters. For our high--significance clusters, Prospector  yields less massive solutions (about half the mass we derive with our approach), with similar ages (median age=6~Myr) but a broader range 5-15~Myr and larger uncertainties ($\Delta Log(Age)\sim$0.8), and similar attenuation values. Only two of the high--significance clusters (\# 7 and \# 18) are found by Prospector to be older, $\sim$15~Myr, and less attenuated, E(B$-$V)$\sim$1.3--1.4~mag, than what we find with our default approach. 
We conclude  that a different fitting approach from our adopted one does not yield a sample of very young sources; the clusters are still relatively old and with significant foreground extinction. 

One remaining source of bias in our age determinations is absorption of LyC photons by dust \citep{Dopita+2003}. In this case, the ionizing photons produced by the massive stars are directly absorbed by dust before they can ionize the gas, thus decreasing the number of free electrons and recombination cascades and, as a consequence, depressing the luminosity of nebular lines and continuum. We use the results from the models of \citet{Krumholz+2009}, \citet{Draine+2011} and \citet{Yeh+2013} to guide this part of the analysis. According to these models, for dust to significantly absorb ionizing photons, the hydrogen inside the HII region needs to be highly ionized, because neutral hydrogen has a much larger cross section than dust for ionizing photon absorption. The high ionization increases the radiation pressure, which thus `pushes' the ionized gas into a shell creating a cavity within the HII region. In this configuration, most of the ionizing photon absorption now occurs in the shell, which is also where the neutral gas lives, in turn limiting the effectiveness of the dust absorption. The result is a `floor' of about 50\%--70\% to the maximum number of ionizing photons that can be directly absorbed by dust at solar metallicity, and this number decreases for decreasing  metal abundances \citep{Draine+2011}. A $\sim$4~Myr  old star cluster has an EW(Pa$\beta$) that is $\sim$17 times larger than that of a 6~Myr old star cluster \citep[from Starburst99,][]{Leitherer+1999}. Thus, in order for a 4~Myr old cluster to `mimick' a 6~Myr old cluster through our SED fitting approach, dust would need to directly absorb $\sim$94\% of the ionizing photons and suppress the Pa$\beta$ line emission accordingly. This is a much larger fraction than the `floor' discussed above, which enables us to disfavor direct absorption of LyC photons by dust as a driver for the best--fit ages.

\subsection{Comparison with the Young, Low--Dust Cluster Population}

The LEGUS project \citep{Calzetti+2015a} isolated star cluster candidates in this galaxy, as well as other galaxies, which were then visually  inspected by LEGUS team members and classified according to four categories: 1,2,3 for star clusters or compact associations and 4 for everything else (contaminant, interlopers, stars with diffuse halos, etc.). Details 
on the selection and visual classification of the star cluster candidates are given in \citet{Adamo+2017}. We summarize here the characteristics of the selection and identification that are relevant to this work. The LEGUS observations covered the five bands  NUV,U,B,V,I (WFC3/F275W, F336W, F438W, F555W and F814W; where possible, archival ACS/F435W, F555W, F814W images were used), and cluster candidates were selected from a white--light image obtained from combining the images in all five bands \citep{Calzetti+2015a}. The automatically--selected star clusters were then visually inspected when brighter than M$_{V, Vega}$=$-$6~mag, which corresponds to a mass of a few hundred M$_{\odot}$. The magnitude cut was imposed after aperture correction, using an average value for the latter \citep{Adamo+2017}.  \citet{Whitmore+2020} added the H$\alpha$ filter to expand the parameter space for selecting star clusters and expanded the existing catalog with visually--selected clusters slightly fainter than the $-$6~mag limit; these authors produced a final catalog of 594 star clusters  in the 1+2+3 classes. The catalog expansion of \citet{Whitmore+2020} has no impact on our analysis since we limit our comparisons to young clusters more massive than $\sim$3,000~M$_{\odot}$, which are also brighter than M$_{V,Vega}$=$-$6~mag. 

As detailed in \citet{Adamo+2017}, the SEDs of all cluster candidates detected in at least four of the five bands were fit with the  same SSP models used in this work, using a similar treatment of the dust attenuation. The requirement that the clusters have to be detected in four out of five bands means that either the NUV (F275W) or the U (F336W) are included in the fits, which imposes a maximum limit on the amount of attenuation in each cluster since the clusters need to be detected at blue wavelengths. The analysis performed by \citet{Whitmore+2020} on the LEGUS star cluster population confirms the maximum observed E(B$-$V)$\lesssim$1~mag. 

\begin{figure}
\plotone{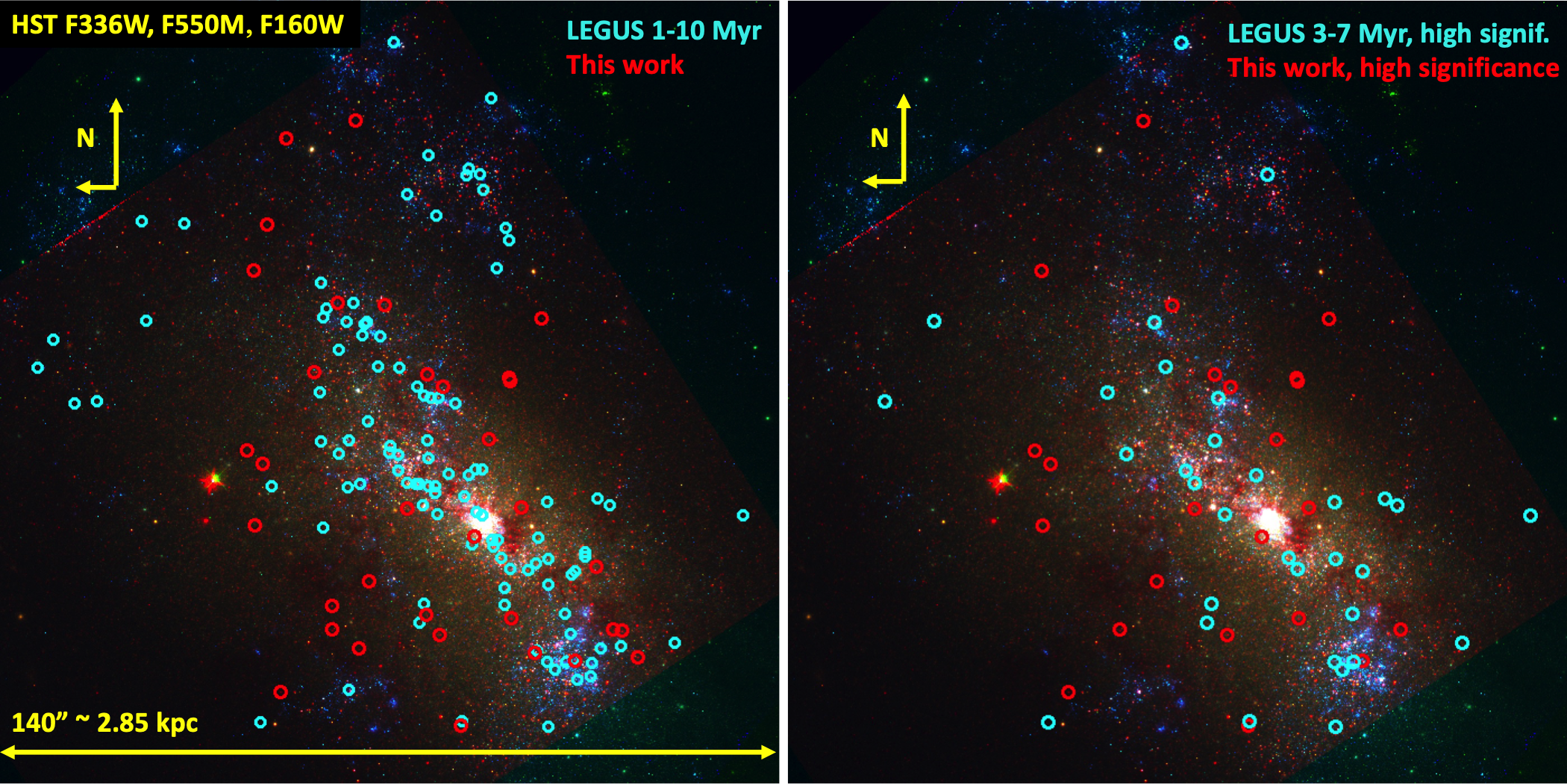}
\caption{(Left:) The location of the LEGUS star clusters with age 1--10~Myr and mass$\ge$3,000~M$_{\odot}$ (cyan circles) and of our dust--buried sources (red circles) is shown on a three--color composite using F336W (blue), F550M (green), and F160W (red). Only  the LEGUS sources located within the WFC3/IR footprint are shown here. (Right:) The same as the left panel, but restricting both types of sources to the high--significance ones: $\chi^2_{red}\le$10 for the LEGUS sources \citep[cyan circles][]{Adamo+2017} and $\chi^2_{red}\le$6 for the dust --buried sources (red circles). The LEGUS sources are further restricted to the age range 3--7~Myr to better match the age range of the dust--buried sources.}
 \label{fig:legus_map}
\end{figure}

The physical parameters, age, mass and E(B$-$V), provided in the LEGUS catalogs are from the Yggdrasil SSPs attenuated by the Milky Way extinction \citep{Fitzpatrick+2019} and the starburst attenuation curves, only. Of the different `flavors' of catalogs available, we use those with aperture corrections derived from isolated clusters measured in the images \citep{Adamo+2017}, which matches our approach to aperture photometry correction. We update the LEGUS masses to account for the slight difference in adopted distance for NGC\,4449 (3.9~Mpc for LEGUS versus 4.2~Mpc here). We select all clusters (classes 1, 2, and 3) with age$\le$10~Myr and mass$\ge$3,000~M$_{\odot}$, detected in at  least four filters. The are 56 such sources from the Milky Way extinction catalog and 127 from the starburst attenuation  catalog, 43 and 106 of which, respectively, are within the footprint of the WFC3/IR images. The difference in source numbers between the two cases is easily understood in terms of the difference in shape between the two curves in the NUV \citep[e.g.][]{Calzetti+1994}. The Milky Way extinction curve has a strong feature at 0.2175~$\mu$m while the starburst curve has none, and the wings of this feature enter in the F275W filter. Thus, if the dust attenuation in NGC\,4449 does  not have intrinsically strong absorption at 0.2175~$\mu$m, the application of the Milky Way extinction to red SEDs will inevitably yield old ages and low extinctions to `force' the fit of the F275W filter. Conversely, the application of the starburst curve enables a wider range of ages and attenuations. We retain only the catalog which uses the starburst attenuation curve to derive the physical parameters, as this case is likely to be more representative of exposed or partially exposed clusters. The distribution of the LEGUS sources relative to the dust--buried sources from this work is shown in Figure~\ref{fig:legus_map}, left.

\begin{figure}
\plottwo{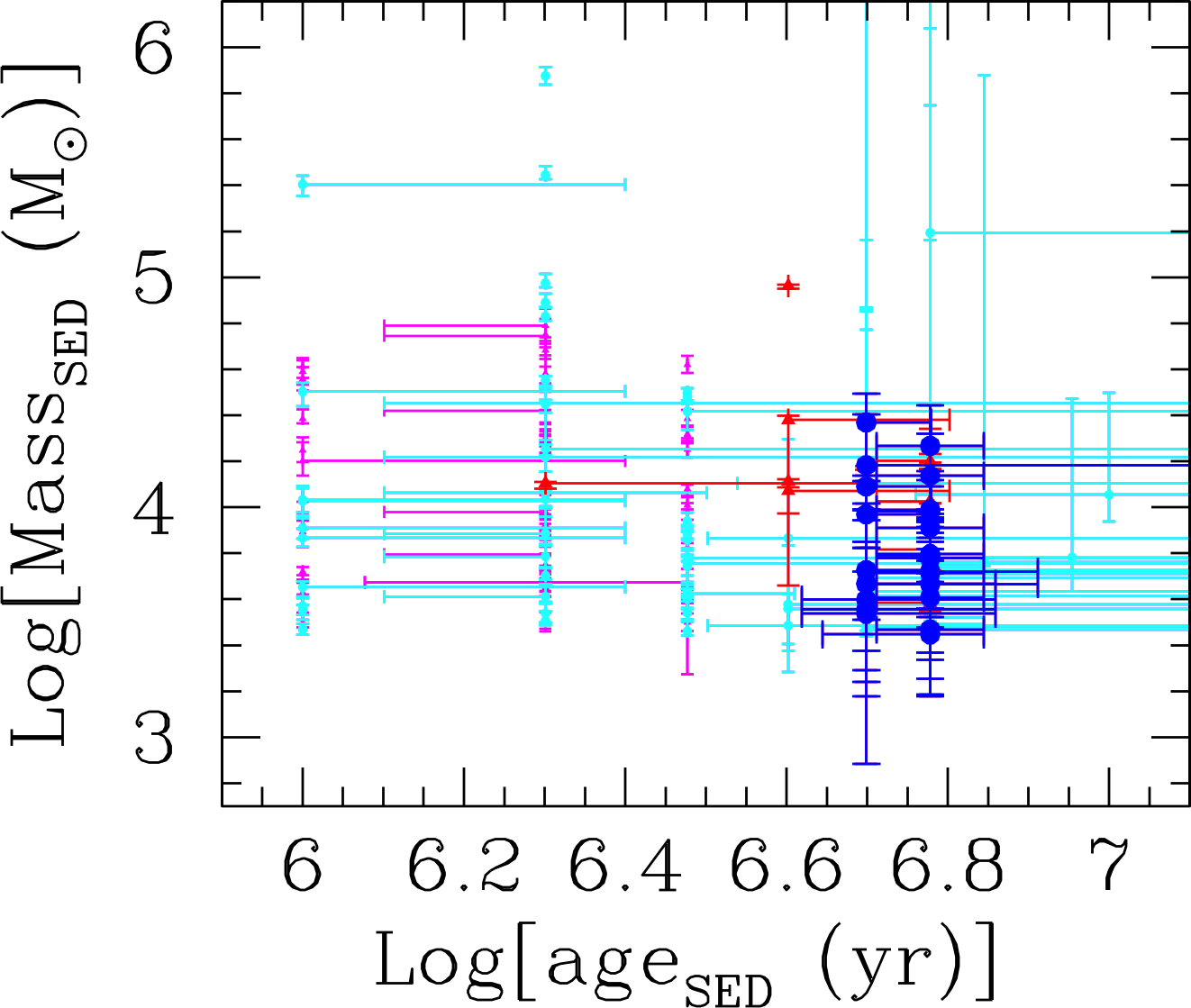}{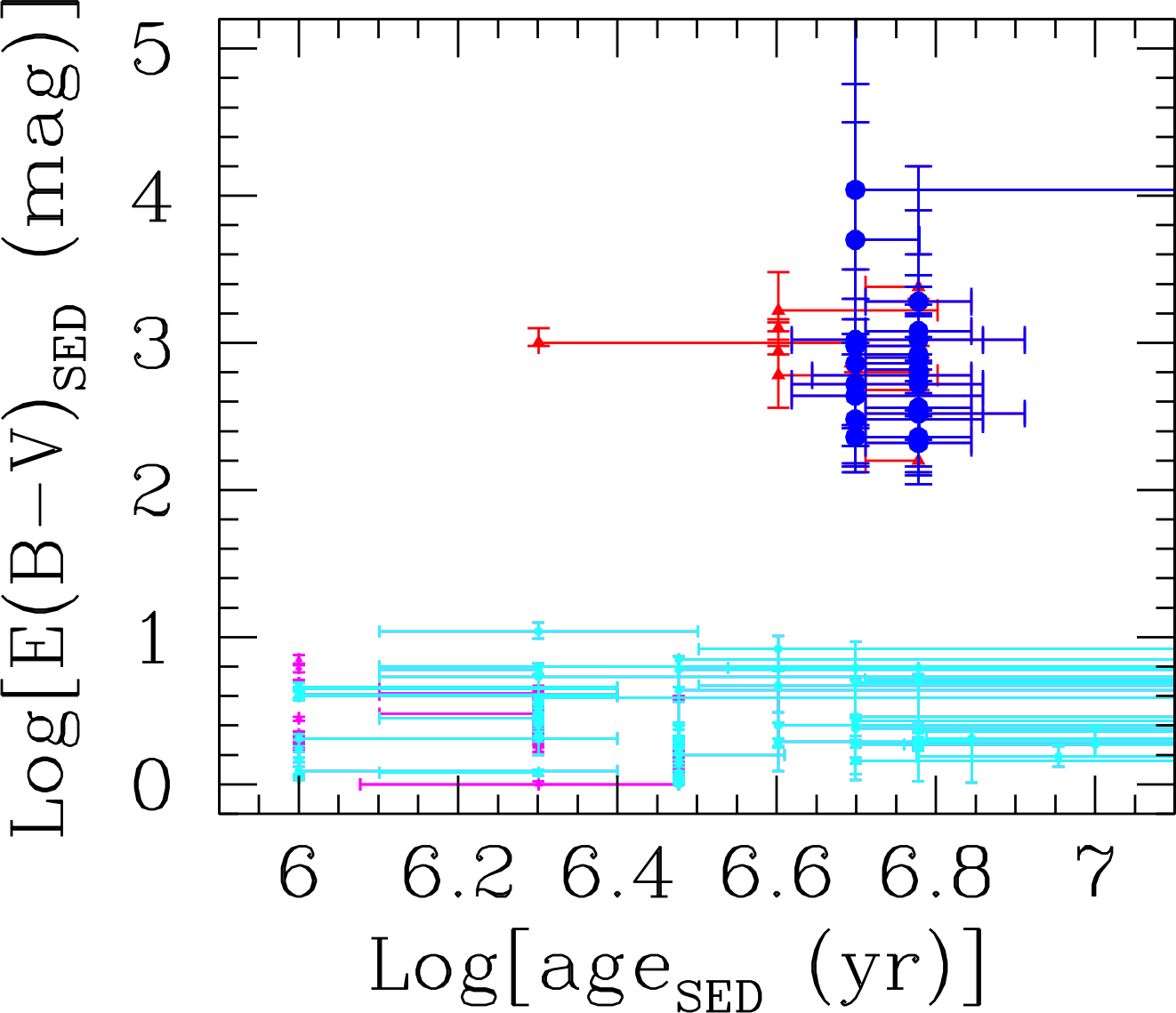}
\caption{The stellar mass (left) and the color excess (right) as a function of the age derived from the SED fits, with 1~$\sigma$ uncertainties, for our sources (blue and red symbols) and for the LEGUS sources (cyan and magenta symbols). As in previous figures, blue symbols indicate sources with $\chi^2_{red}\le6$ from the SED fits, while red symbols are for sources with larger values of $\chi^2_{red}$. For the LEGUS sources, the cyan symbols are for sources with $\chi^2_{red}\le10$ and magenta symbols for sources with larger values of $\chi^2_{red}$.}
\label{fig:legus}
\end{figure}

Figure~\ref{fig:legus} shows the mass--versus--age and the E(B$-$V)--versus--age plots of our sources (blue and red symbols as before) overlayd on top of the same parameters for the 106 LEGUS sources (cyan and magenta symbols). The cyan symbols identify the 53 sources with $\chi^2_{red}\le10$, thus those that have high--fidelity SED fits \citep{Adamo+2010}. Of these, 32 (16) have ages between 3 (4) and 7 Myr, to be compared with  our 23 high--fidelity sources (Figure~\ref{fig:legus_map}, right). The two samples overlap in mass range, although the LEGUS sample extends to higher mass values, almost 10$^6$~M$_{\odot}$, than our sample. This is due to the selection function for our sources, which imposes low--significance or non--detections in H$\alpha$, thus excluding high mass (brighter) sources. However, the two samples occupy different spatial locations within the galaxy, with the LEGUS high--fidelity, 3-7~Myr sources located mainly along the bright  ridge of star  formation and the dust--buried sources distributed more homogeneously throughout the FoV(Figure~\ref{fig:legus_map}, right). All selected  LEGUS sources have E(B$-$V)$\lesssim$1~mag, as already discussed. Thus our sources and the LEGUS sources occupy two distinct loci in the E(B$-$V) parameter space, as can be expected since the two samples have been selected with drastically different properties: UV and optically faint for our sources and UV detected for the LEGUS sources. Further comparisons would need to take into account the differences between the two samples, including differences in the SED fitting approach and wavelength range; however we note that, within a similar age range, they contain comparable numbers of sources. This indicates that our sample is not simply a collection of rare outliers, but represents a population of dust--buried clusters with a comparable census as the low--dust clusters.

\subsection{The Location of the Dust--Buried Clusters}

Given the dusty nature of our sources and their location in correspondence of dust lanes/clouds, we  consider the possibility that our selection criteria isolate 
star clusters that have emerged on {\em the far side} of the galaxy or of the clouds where they formed. We present two arguments to exclude this possibility: one based on the spatial distribution of the dust--buried and low--dust clusters and one based on the characteristics of the two populations.

We first evaluate whether the dust--buried clusters are located behind the galaxy's mid--plane dust. Measurements of the typical column densities of mid--plane diffuse dust indicate values of A$_V\lesssim$1--1.5~mag in SDSS spiral galaxies \citep{Holwerda+2007}. Even accounting for the inclination of NGC\,4449 (Table~\ref{table:galaxy}), we obtain a line--of--sight column density of A$_V\sim$2.7--4.0~mag for the mid--plane dust, which is  significantly lower than the dust column densities we derive (A$_V >$6~mag). In addition, several of our sources are located in the outskirts of the brightest areas of dust emission (Figure~\ref{fig:individual_cutouts}), suggesting that they are at the margins of actively star forming, and possibly dense, regions. Both pieces of evidence suggest that the clusters in our sample are affected by local dust absorption, as can be caused by the colocation with natal clouds.  

We use the census of dust--buried clusters to evaluate the possibility  that  these are clusters that have emerged on the {\em far side} of molecular clouds. Statistical considerations  expect comparable numbers of  star clusters to emerge from both the far side and the near side of clouds. Observationally, we expect to observe a larger number of clusters emerging from the near side  of the cloud than the far side, for two reasons: (1) some of the {\em far side} clusters will be dimmed below detectability by the foreground dust; (2)  existing models predict that natal molecular clouds get destroyed by stellar feedback, thus freeing star clusters of their  dust. 
The `dark clouds', i.e., regions that are well detected both in the H--band filter and in the 8~$\mu$m dust map but are weakly emitting in the V--band, occupy about 20\% of the total star--forming area within the WFC3/IR FoV. Based on areal coverage arguments  alone, we expect about 5 times more low--dust sources than dust--buried ones in the 3--7~Myr age range.  We detect only 33\% more high--significance low--dust clusters than dust--buried ones. The addition of undetectable, deeply  dust--buried clusters is unlikely to change statistics significantly: the area of  the dark clouds that have A$_V>$10~mag, has low filling factor  \citep[$<$10\%,][]{Lada+2009}. For this estimate we assume that young, low--dust clusters that are far away from any identified dark cloud were born in clouds that  have since been destroyed. In a second estimate, we limit our calculation to the dark clouds hosting or in proximity of our 23 high--significance dust--buried clusters. In  this case we expect  about equal numbers  of dust--buried and low--dust clusters, and possibly larger numbers of low--dust clusters if some dust--buried ones are lost from census because they are so dimmed by dust to be undetectable.  We count 13 low--dust clusters in the 3--7~Myr age range, i.e., about (or less than) 1/2 of what would be expected based on migration considerations alone. 

The spatial arguments above, however, do not leverage the main characteristic of the dust--buried clusters: they show compact emission in the Pa$\beta$ line, with EWs in the range 25~\AA--125~\AA\ (Table~\ref{tab:photometry}). We measure the Pa$\beta$ emission in the 53 low--dust, 1--10~Myr old clusters of the LEGUS sample, adopting the same photometric aperture of 0$^{\prime\prime}$12 radius used for our dust--buried sources. Prior to performing aperture photometry, we carefully determine the centroids of the low--dust clusters using the higher resolution optical images. We find that only four low--dust clusters have detectable line emission within those apertures, with EW(Pa$\beta$)$\simeq$210~\AA, 165~\AA, 4~\AA, and 2~\AA. All four clusters are younger than 3~Myr. The remaining 49 low--dust clusters are consistent with having zero Pa$\beta$ line emission in the small photometric aperture, including all clusters with age  $\ge$3~Myr. Thus, {\em none of the low--dust clusters has values of the EW(Pa$\beta$) matching those of our dust--buried population}. A visual inspection of the continuum--subtracted Pa$\beta$ and H$\alpha$ images confirms that only 4 sources (8\% of the total) have centrally--peaked nebular emission. Of the remaining 49, 12 are located in areas with no nebular emission within a radius of $\sim$40~pc, while the other  37 are surrounded by nebular emission shells, broken or whole, and filaments. Increasing the aperture to 0$^{\prime\prime}$20 (5~pixels) radius only adds two clusters with positive Pa$\beta$ line detection, one younger than 3~Myr with EW(Pa$\beta$)$\simeq$1~\AA\ and one 5~Myr old with EW(Pa$\beta$)$\simeq$5~\AA. 

As argued above, if the dust--buried clusters were truly low--dust clusters that have emerged on the far side of their natal clouds, we would expect roughly equal numbers of clusters with similar properties which have emerged on the near side of the clouds. Alas, we find no low--dust clusters with compact Pa$\beta$ line emission in the age range 3--7~Myr and none with EW(Pa$\beta$) in the range 25~\AA--125~\AA\ irrespective of age; conversely, we have 23 dust--buried clusters with such characteristics. Under the assumption of a Poisson distribution for the spatial distribution of clusters, the probability of finding zero clusters on one side and 23 on the other side of clouds is p(0)$\sim$1$\times$10$^{-10}$; even under the generous assumption that all four clusters younger than 3~Myr and with detected Pa$\beta$ emission could resemble our dust--buried population, the probability p($\le$4)$\sim$1.4$\times$10$^{-6}$. The implication is that the low--dust and the dust--buried clusters represent two morphologically distinct populations. The low--dust, young star clusters are consistent with early gas clearing \citep[e.g.,][]{Krumholz+2019, Dinnbier+2020}, while the dust--buried clusters appear to have retained their gas as would occur if the gas were prevented from leaving the area of the clusters (see next section). The two populations could actually represent the two extremes of a continuum of properties, ranging from clusters that undergo early gas clearing to clusters that are resistant to gas clearing, possibly due to variations in local conditions.

Our estimates, therefore, indicate that a large fraction of the dust--buried clusters are truly embedded or  partially embedded in their natal clouds and are unlikely to be sources that have emerged on the far side of  the galaxy or the clouds.

\subsection{Implications for the Emergence Timescale(s) of Star Clusters}

Pre--SN feedback processes that are likely to be important for young star clusters in our mass and size range (3,000--25,000~M$_{\odot}$, $\approx$1--3~pc, v$_{esc}\sim$7~km~s$^{-1}$) include photoionization, direct radiation pressure and stellar winds \citep{Pellegrini+2011, Dale+2012, Krause+2013, Krumholz+2019}, all of which have timescales of a few Myr, and should help clear the surrounding medium before the first supernova explosion occurs at around $\sim$4~Myr \citep{Leitherer+2014}. The presence of a population of young ($\sim$1--6~Myr) star clusters with low extinction values, as identified by the LEGUS collaboration \citep{Whitmore+2020}, aligns with this framework of early gas dispersal from around the clusters.

Conversely, the population of dust--buried  star  clusters identified in this  work does not support that general picture, and requires a different explanation. These clusters are old enough, according to their SED--derived ages of $\sim$5--6~Myr, that they should have experienced both pre--SN and supernova feedback. We note that our results for the ages are robust against choices of the attenuation/extinction prescription. Supernovae may have 
not occurred yet if, due to incomplete sampling of the IMF, the clusters' most massive stars are $\lesssim$30~M$_{\odot}$ \citep[i.e., lifetimes $\gtrsim$6~Myr,][]{Castelli+2003}. \citet{Chevance+2022} derive a range of supernova timescales, $\sim$4--8~Myr for our cluster mass range, by including stochastic IMF sampling. These clusters, however, still contain sufficiently massive stars that they  
should have experienced pre--SN feedback \citep{Leitherer+2014} and performed some clearing of the surrounding medium; alas, they appear not to  have done so. Using the measured A$_V\sim$6.5--11~mag range to estimate the gas column density in front of the clusters, we obtain N(H)$\sim$(3.4--5.7)$\times$10$^{22}$~cm$^{-2}$ \citep{Bohlin+1978, Zhu+2017}, after including that NGC\,4449  has 40\% solar metallicity and thus larger gas densities at given A$_V$ than a solar metallicity  source \citep{RemyRuyer+2014}. If the  column of gas and dust in  front of the dust--buried clusters is $\sim$20~pc in depth  (i.e., 1/2  the radius of a typical Milky Way's molecular cloud), the resulting density is: n(H)=550--920~cm$^{-3}$. If the same density fully surrounds the star cluster, it is likely that the ejection and expansion of its gas in response to feedback has been stalled, as we discuss in the next paragraphs \citep[e.g.,][]{Smith+2006, Silich+2007, Westmoquette+2014, Zamora+2019}. 

The measured gas column densities correspond to an external pressure of P$_{ext}$/$k_B$=3.9319$\times$10$^{-39}$~$N(H)^2$=0.5--1.3$\times$10$^7$~K~cm$^{-3}$, where N(H) is in units of cm$^{-2}$ (the numerical constant is not adimensional); we neglect  the contribution of  the stellar component to the pressure and assume a 1.36 multiplying factor to include He \citep{Elmegreen1989}. For  the HII regions surrounding the clusters, we calculate the pressure  contribution from photoionization, direct radiation pressure, and stellar winds. We do not include dust--reprocessed radiation pressure because our clusters have stellar surface densities $\lesssim$10$^4$~M$_{\odot}$~pc$^{-2}$, i.e. they are about an order of magnitude below the minimum stellar surface density for dust--reprocessed radiation pressure to significantly contribute to feedback \citep{Krumholz+2019}.  We consider the two extreme cases of a 25,000~M$_{\odot}$, 4~Myr star cluster with size 3~pc and a 3,000~M$_{\odot}$, 6~Myr star cluster with size 1~pc to bracket the observed range of the dust--buried sources. For these two extreme cases, we use the Starburst99 ionizing photon fluxes to derive electron densities of 590~cm$^{-3}$ and 350~cm$^{-3}$, respectively, using a Str\"omgren sphere approximation. The electron densities in the HII regions  are slightly lower than those derived for the gas surrounding them.  The photoionization and direct radiation pressures are calculated using the formulae published by \citet{Lopez+2014}. For the photoionization, the get P$_{ph}$/$k_B$=0.7--1.2$\times$10$^7$~K~cm$^{-3}$, while for the radiation pressure P$_{rad}$/$k_B$=2.4--4.4$\times$10$^7$~K~cm$^{-3}$. We derive the stellar wind pressure using the formula of \citet{Weaver+1977} as formulated in \citet{Smith+2006}. We obtain that the bubble pressure is P$_{wind}$/$k_B$=3.1--17.3$\times$10$^7$~K~cm$^{-3}$ for 100\% thermal efficiency and 0.7--3.7$\times$10$^7$~K~cm$^{-3}$ for 10\% thermal efficiency. Each of these pressure terms is comparable or slightly  larger (but only by a factor of a few) than the external pressure, supporting the inference above that the HII regions are not expanding and are likely to have stalled, due possibly to radiative cooling \citep{Silich+2007}.  

\citet{Smith+2006} and \citet{Westmoquette+2014} analyze three clusters in the nearby starburst  galaxy  M\,82 that appear to have encountered a fate similar to our dust--buried clusters. The spectroscopic ages of these three clusters, 4.5--6.4~Myr, are similar to  those of the dust--buried star clusters found in this work. They have large  attenuations, E(B$-$V)$\simeq$1.4--1.9~mag, but significantly higher masses than our clusters, by a factor almost 100. Yet, these star clusters are also found to have stalled due to the high  pressure  environment in the central region of  M\,82. For comparison, \citet{Smith+2006} and \citet{Westmoquette+2014} determine values of the external pressure comparable to those we find. Similarly, \citet{DellaBruna+2022} find that HII regions in the center of the starburst galaxy M\,83 are stalled because the ambient pressure is larger than the pre--SN pressure.  

For completeness, we consider the possibility that, because our dust--buried clusters are young, relatively massive, and compact, gas removal may be prevented by self--gravity \citep{Krause+2016, Krause+2020}. \citet{Krause+2016} derive a relation between the compactness of a star cluster, defined as the Mass/(Half--Mass Radius), and the star formation efficiency. Clusters that form with sufficiently high efficiency at fixed compactness are effective at removing gas from their immediate surroundings.  If we adopt our I--band half--light radius as a tracer of the half--mass radius, the compactness of our clusters spans the range 0.6--1.7$\times$10$^4$~M$_{\odot}$~pc$^{-1}$. For this range, self--gravity does not inhibit gas expulsion when star formation efficiencies are above 20\%--30\%  \citep{Krause+2016}. Although we do not have values for the star formation efficiency of our dust--buried star clusters, other estimates suggest this to be about 30\% when evaluated at the sizes of clumps and clusters \citep[e.g.,][]{Lada+2003, Calzetti+2015b, Krumholz+2019} and could increase with increasing stellar mass up to $\sim$60\% at 10$^6$~M$_{\odot}$ \citep{Turner+2015}. If these efficiencies are applicable to our dust--buried clusters, the impact of self--gravity at preventing gas clearance is expected to be secondary relative to the other effects discussed above.

We note that  the population we identify is likely to be a lower limit  to the full dusty population in the 5--6~Myr age range. Our sources  are explicitly selected to have  hydrogen recombination  line emission. As we are  at the boundary where stochastic sampling of the stellar IMF begins to become important, we expect larger numbers of dusty, 5--6~Myr star clusters to have been missed by our search due to the absence  of line emission \citep{Fumagalli+2011}. The addition of these clusters would only increase the dusty population and its implications for the lack of effectiveness of pre--SN feedback.  

\section{Summary and Conclusions} \label{sec:conclusions}

Using multi--wavelength {\em HST} imaging data which include NUV, optical and near--IR, including narrow--band filters centered at the hydrogen line emission of H$\alpha$ and Pa$\beta$, we isolate a population of highly extincted star clusters, with A$_V\sim$6--11~mag and ages $\sim$5--6~Myr. These clusters  are numerous enough to be comparable to the low--dust cluster population identified using NUV--optical images in the same age and mass ranges \citep{Adamo+2017}.  Despite the similarity in numbers, however, the two populations possess morphologically distinct nebular emission properties: the low--dust clusters are consistent with early gas clearing, while the highly--extincted clusters have retained their gas and show compact Pa$\beta$ emission. We use this difference, together with other arguments, to exclude spurious effects, such as that the highly extincted clusters could have emerged behind the clouds where they formed, although we cannot exclude that at least some in our population are of this nature. The highly extincted clusters are, therefore, likely to be buried or partially buried in their natal clouds. 

Stochastic sampling of the stellar Initial Mass Function could have some effect on the derived ages, masses and extinctions, although we present arguments, based on the spatial distribution of the dust--buried clusters, that exclude a large impact of stochasticity on our results. We also exclude direct absorption by dust of ionizing 
photons as a major driver for the relatively old ages we derive from SED fits. The dearth of clusters younger than $\sim$4~Myr in our sample can be understood in terms of our selection criteria, which require our sources to be simultaneously detected in all near--IR bands and undetected or marginally detected in all NUV and optical bands, including detection in the Pa$\beta$ emission line and non/marginal detection in H$\alpha$. 

The dust--buried clusters are sufficiently old that pre--SN feedback should have already cleared the natal cloud; however, the amount of dust in front of the clusters indicates that pre--SN feedback  has not been effective. Furthermore, the surrounding gas has sufficient high pressure and  density that the HII regions around the clusters are likely to have been stalled. These lines of evidence paint a scenario for which a significant fraction of  clusters in the mass range 0.3--2.5$\times$10$^4$~M$_{\odot}$ do not emerge from their dust clouds in a short timescale, but remain within the cloud for at least 6~Myr. Such finding challenges models that require pre--SN feedback to be highly effective at clearing the natal clouds in order to keep the star formation efficiency low and suggests that the pressure from the natal cloud plays an important role in determining the timescale for such clearing. Upcoming JWST observations of this and other galaxies, especially targeting the infrared emission of hydrogen recombination lines, will be key for providing a more complete picture of the nature of dusty star cluster populations.

\begin{acknowledgments}

The authors thank an anonymous reviewer whose comments have helped strengthen the case presented in the manuscript.

Based on observations made with the NASA/ESA Hubble Space Telescope, obtained  at the Space Telescope Science Institute, which is operated by the 
Association of Universities for Research in Astronomy, Inc., under NASA contract NAS 5--26555. These observations are associated with program \# 15330. 
Support for program \# 15330 was provided by NASA through a grant from the Space Telescope Science Institute.

Based also on observations made with the NASA/ESA Hubble Space Telescope, and obtained from the Hubble Legacy Archive, 
which is a collaboration between the Space Telescope Science Institute (STScI/NASA), the Space Telescope European Coordinating 
Facility (ST-ECF/ESA) and the Canadian Astronomy Data Centre (CADC/NRC/CSA).

Most of the data presented in this paper were obtained from the Mikulski Archive for Space Telescopes (MAST)
at the Space Telescope Science Institute. The specific observations analyzed can be accessed via 
\dataset[10.17909/8k2f-7205]{https://doi.org/DOI}.

This research has made use of the NASA/IPAC Extragalactic Database (NED) which is operated by the Jet
Propulsion Laboratory, California Institute of Technology, under contract with the National Aeronautics and Space
Administration.

MRK acknowledges support from the Australian Research Council through Laureate Fellowship FL220100020.

MM acknowledges support from the Swedish Research Council (Vetenskapsr\aa det project grant 2019-00502). 

\end{acknowledgments}

\vspace{5mm}
\facilities{Hubble Space Telescope (WFC3, ACS)}
\software{Drizzlepac \citep[][and the STSCI Development Team]{Gonzaga+2012}, IRAF \citep{Tody1986, Tody1993}, SAOImage DS9 \citep{Joye+2003}, Fortran, Prospector \citep{Johnson+2021}}

\begin{deluxetable}{lrrrrrrrrrrrrr}
\tablecolumns{14}
\rotate
\tabletypesize{\tiny}
\tablecaption{Source Location and Photometry\label{tab:photometry}}
\tablewidth{0pt}
\tablehead{
\colhead{ID} & \colhead{RA(2000),DEC(2000)}  & \colhead{F275W} & \colhead{F336W} & \colhead{F435W} & \colhead{F555W} 
& \colhead{F550M} & \colhead{F658N} & \colhead{F814W} & \colhead{F110W} & \colhead{F128N} & \colhead{F160W}  & \colhead{L(Pa$\beta$)}  & \colhead{EW(Pa$\beta$)} 
\\
\colhead{(1)} & \colhead{(2)} & \colhead{(3)} & \colhead{(4)} & \colhead{(5)}  
& \colhead{(6)} & \colhead{(7)} & \colhead{(8)} & \colhead{(9)} & \colhead{(10)} & \colhead{(11)} & \colhead{(12)} & \colhead{(13)}  & \colhead{(14)} 
\\
}
\startdata
\hline
  1 &12:28:13.6083, +44:05:16.888 & $<$32.326 & $<$32.054 & $<$32.090 & $<$32.084 & $<$32.169 &32.664$\pm$0.413 &33.268$\pm$0.104 &33.736$\pm$0.032 &33.868$\pm$0.076 &33.706$\pm$0.039 &35.528$\pm$0.078 &1.806$\pm$0.086\\
  2 &12:28:13.1596, +44:05:13.457 & $<$32.115 & $<$31.843 & $<$31.879 & $<$31.872 & $<$31.958 &32.713$\pm$0.370 &33.430$\pm$0.086 &33.570$\pm$0.039 &33.683$\pm$0.095 &33.443$\pm$0.053 &35.365$\pm$0.097 &1.838$\pm$0.109\\
  3 &12:28:13.6158, +44:05:21.239 & $<$32.012 & $<$31.739 & $<$31.776 & $<$31.769 & $<$31.854 &32.449$\pm$0.516 &33.152$\pm$0.119 &33.872$\pm$0.028 &33.931$\pm$0.071 &33.845$\pm$0.034 &35.306$\pm$0.073 &1.445$\pm$0.080\\
  4 &12:28:13.0039, +44:05:25.485 & $<$31.966 & $<$31.694 & $<$31.730 & $<$31.724 & $<$31.809 &32.449$\pm$0.569 &32.988$\pm$0.143 &33.359$\pm$0.049 &33.564$\pm$0.108 &33.472$\pm$0.051 &35.276$\pm$0.111 &1.881$\pm$0.123\\
  5 &12:28:12.0411, +44:05:19.530 & $<$32.228 & $<$31.955 & $<$31.991 & $<$31.985 & $<$32.070 &32.914$\pm$0.291 &33.134$\pm$0.121 &33.824$\pm$0.030 &33.876$\pm$0.076 &33.696$\pm$0.040 &35.363$\pm$0.078 &1.581$\pm$0.086\\
  6 &12:28:11.8231, +44:05:16.022 & $<$32.472 & $<$32.200 & $<$32.236 & $<$32.229 & $<$32.313 &32.989$\pm$0.271 &33.449$\pm$0.085 &33.783$\pm$0.031 &33.873$\pm$0.076 &33.755$\pm$0.037 &35.396$\pm$0.078 &1.625$\pm$0.086\\
  7 &12:28:11.2588, +44:05:33.491 & $<$32.852 & $<$32.578 & $<$32.616 & $<$32.609 & $<$32.694 &33.061$\pm$0.275 &33.549$\pm$0.076 &33.721$\pm$0.033 &33.815$\pm$0.081 &33.689$\pm$0.040 &35.358$\pm$0.083 &1.650$\pm$0.091\\
  8 &12:28:11.0066, +44:05:50.971 & $<$32.933 & $<$32.660 & $<$32.696 & $<$32.689 & $<$32.775 &33.393$\pm$0.162 &33.200$\pm$0.117 &33.557$\pm$0.041 &33.727$\pm$0.090 &33.495$\pm$0.051 &35.488$\pm$0.093 &1.956$\pm$0.104\\
  9 &12:28:10.4670, +44:05:38.820 & $<$32.393 & $<$32.119 & $<$32.157 & $<$32.150 & $<$32.235 &32.464$\pm$0.584 &33.050$\pm$0.134 &33.577$\pm$0.039 &33.701$\pm$0.093 &33.575$\pm$0.046 &35.308$\pm$0.095 &1.734$\pm$0.105\\
 10 &12:28:12.0334, +44:06:02.626 & $<$32.366 & $<$32.092 & $<$32.129 & $<$32.123 & $<$32.208 &32.848$\pm$0.321 &33.092$\pm$0.127 &33.753$\pm$0.032 &33.875$\pm$0.076 &33.737$\pm$0.038 &35.489$\pm$0.078 &1.744$\pm$0.086\\
 11 &12:28:08.5375, +44:05:12.052 & $<$32.630 & $<$32.356 & $<$32.394 & $<$32.387 & $<$32.472 &33.433$\pm$0.155 &33.409$\pm$0.089 &33.940$\pm$0.026 &34.014$\pm$0.065 &33.832$\pm$0.034 &35.563$\pm$0.066 &1.660$\pm$0.074\\
 12 &12:28:08.8027, +44:05:16.732 & $<$32.687 & $<$32.415 & $<$32.451 & $<$32.444 & $<$32.529 &32.638$\pm$0.503 &33.451$\pm$0.085 &33.948$\pm$0.026 &34.017$\pm$0.065 &33.834$\pm$0.034 &35.558$\pm$0.066 &1.648$\pm$0.074\\
 13 &12:28:08.9467, +44:05:16.868 & $<$32.439 & $<$32.168 & $<$32.203 & $<$32.197 & $<$32.282 &32.449$\pm$0.880 &32.768$\pm$0.186 &33.267$\pm$0.055 &33.576$\pm$0.107 &33.623$\pm$0.043 &35.261$\pm$0.110 &1.843$\pm$0.123\\
 14 &12:28:09.5793, +44:05:11.430 & $<$33.203 & $<$32.932 & $<$32.967 & $<$32.962 & $<$33.047 &33.201$\pm$0.962 &33.435$\pm$0.093 &34.032$\pm$0.025 &34.129$\pm$0.057 &33.988$\pm$0.029 &35.694$\pm$0.059 &1.680$\pm$0.065\\
 15 &12:28:10.2393, +44:05:12.852 & $<$33.171 & $<$32.897 & $<$32.935 & $<$32.928 & $<$33.013 &34.008$\pm$0.164 &34.327$\pm$0.032 &34.507$\pm$0.015 &34.532$\pm$0.036 &34.365$\pm$0.019 &35.902$\pm$0.037 &1.439$\pm$0.041\\
 16 &12:28:10.6461, +44:05:18.892 & $<$32.583 & $<$32.310 & $<$32.347 & $<$32.340 & $<$32.424 &32.449$\pm$0.753 &32.924$\pm$0.156 &33.852$\pm$0.029 &33.999$\pm$0.066 &33.938$\pm$0.030 &35.580$\pm$0.068 &1.700$\pm$0.075\\
 17 &12:28:09.2287, +44:05:28.172 & $<$32.472 & $<$32.200 & $<$32.236 & $<$32.229 & $<$32.314 &32.449$\pm$0.576 &32.707$\pm$0.200 &33.665$\pm$0.035 &33.742$\pm$0.088 &33.479$\pm$0.051 &35.370$\pm$0.091 &1.764$\pm$0.101\\
 18 &12:28:12.3614, +44:05:38.604 & $<$32.820 & $<$32.548 & $<$32.584 & $<$32.576 & $<$32.661 &32.449$\pm$0.678 &33.493$\pm$0.081 &33.660$\pm$0.036 &33.804$\pm$0.082 &33.605$\pm$0.045 &35.508$\pm$0.085 &1.871$\pm$0.094\\
 19 &12:28:13.5209, +44:06:15.261 & $<$32.853 & $<$32.579 & $<$32.617 & $<$32.610 & $<$32.695 &33.262$\pm$0.197 &33.630$\pm$0.070 &33.608$\pm$0.038 &33.748$\pm$0.088 &33.489$\pm$0.051 &35.483$\pm$0.090 &1.917$\pm$0.102\\
 20 &12:28:12.7311, +44:06:14.889 & $<$32.531 & $<$32.257 & $<$32.295 & $<$32.287 & $<$32.373 &32.714$\pm$0.418 &33.153$\pm$0.119 &33.535$\pm$0.041 &33.690$\pm$0.094 &33.517$\pm$0.049 &35.395$\pm$0.096 &1.870$\pm$0.107\\
 21 &12:28:14.6910, +44:06:29.325 & $<$32.072 & $<$31.799 & $<$31.836 & $<$31.829 & $<$31.914 &32.697$\pm$0.374 &33.431$\pm$0.086 &33.549$\pm$0.040 &33.651$\pm$0.098 &33.530$\pm$0.048 &35.206$\pm$0.100 &1.665$\pm$0.110\\
 22 &12:28:14.9099, +44:06:21.084 & $<$32.417 & $<$32.145 & $<$32.181 & $<$32.174 & $<$32.259 &32.591$\pm$0.452 &32.813$\pm$0.176 &33.508$\pm$0.042 &33.652$\pm$0.098 &33.424$\pm$0.054 &35.374$\pm$0.101 &1.897$\pm$0.112\\
 23 &12:28:14.3714, +44:06:44.803 & $<$32.181 & $<$31.908 & $<$31.945 & $<$31.938 & $<$32.022 &32.967$\pm$0.265 &33.807$\pm$0.056 &33.559$\pm$0.039 &33.723$\pm$0.090 &33.494$\pm$0.050 &35.474$\pm$0.093 &1.942$\pm$0.103\\
24 &12:28:13.2234, +44:06:47.993 & $<$32.096 & $<$31.824 & $<$31.860 & $<$31.854 & $<$31.939 &32.449$\pm$0.552 &33.091$\pm$0.127 &33.270$\pm$0.054 &33.426$\pm$0.127 &33.181$\pm$0.072 &35.180$\pm$0.130 &1.944$\pm$0.146\\
 25 &12:28:13.9051, +44:06:03.009 & $<$32.537 & $<$32.266 & $<$32.301 & $<$32.295 & $<$32.380 &32.449$\pm$0.579 &32.553$\pm$0.243 &33.922$\pm$0.027 &33.965$\pm$0.069 &33.803$\pm$0.035 &35.403$\pm$0.070 &1.520$\pm$0.078\\
 26 &12:28:11.7727, +44:06:00.424 & $<$32.532 & $<$32.258 & $<$32.295 & $<$32.289 & $<$32.374 &32.449$\pm$0.710 &33.160$\pm$0.118 &33.466$\pm$0.044 &33.624$\pm$0.101 &33.382$\pm$0.057 &35.372$\pm$0.104 &1.936$\pm$0.116\\
 27 &12:28:10.6799, +44:06:01.920 & $<$32.188 & $<$31.916 & $<$31.952 & $<$31.945 & $<$32.030 &32.729$\pm$0.368 &33.112$\pm$0.124 &33.619$\pm$0.037 &33.653$\pm$0.098 &33.533$\pm$0.048 &34.979$\pm$0.100 &1.390$\pm$0.109\\
 28 &12:28:10.6474, +44:06:01.332 & $<$32.129 & $<$31.856 & $<$31.892 & $<$31.886 & $<$31.971 &32.449$\pm$0.603 &32.318$\pm$0.311 &33.144$\pm$0.063 &33.380$\pm$0.134 &33.123$\pm$0.077 &35.224$\pm$0.138 &2.095$\pm$0.156\\
 29 &12:28:10.1293, +44:06:12.579 & $<$32.093 & $<$31.821 & $<$31.858 & $<$31.851 & $<$31.936 &32.725$\pm$0.359 &32.608$\pm$0.222 &33.600$\pm$0.038 &33.698$\pm$0.093 &33.615$\pm$0.044 &35.193$\pm$0.095 &1.590$\pm$0.104\\
 30 &12:28:11.4805, +44:04:59.731 & $<$32.173 & $<$31.901 & $<$31.937 & $<$31.930 & $<$32.015 &33.053$\pm$0.243 &32.800$\pm$0.178 &33.646$\pm$0.036 &33.829$\pm$0.080 &33.867$\pm$0.033 &35.339$\pm$0.082 &1.608$\pm$0.090\\
 31 &12:28:14.4625, +44:05:05.845 & $<$32.365 & $<$32.092 & $<$32.129 & $<$32.122 & $<$32.207 &33.134$\pm$0.220 &33.025$\pm$0.137 &33.633$\pm$0.036 &33.789$\pm$0.084 &33.608$\pm$0.044 &35.497$\pm$0.086 &1.876$\pm$0.096\\
 32 &12:28:14.8929, +44:05:35.564 & $<$32.295 & $<$32.022 & $<$32.059 & $<$32.051 & $<$32.137 &32.449$\pm$0.530 &32.820$\pm$0.174 &33.335$\pm$0.051 &33.535$\pm$0.112 &33.268$\pm$0.065 &35.347$\pm$0.116 &2.042$\pm$0.130\\
 33 &12:28:14.7671, +44:05:46.484 & $<$32.590 & $<$32.318 & $<$32.354 & $<$32.348 & $<$32.433 &32.739$\pm$0.378 &33.144$\pm$0.121 &33.415$\pm$0.047 &33.553$\pm$0.110 &33.360$\pm$0.059 &35.243$\pm$0.113 &1.851$\pm$0.126\\
 34 &12:28:15.0205, +44:05:49.044 & $<$31.987 & $<$31.714 & $<$31.751 & $<$31.744 & $<$31.829 &32.539$\pm$0.464 &33.044$\pm$0.134 &33.234$\pm$0.057 &33.380$\pm$0.134 &33.195$\pm$0.071 &35.077$\pm$0.137 &1.862$\pm$0.153\\
\enddata

\tablenotetext{}{(1) The identification number of the source.}
\tablenotetext{}{(2) Right Ascension  and Declination in J2000 coordinates.}
\tablenotetext{}{(3)--(12) Logarithm of the luminosity density of each source in the indicated filter, in units or erg~s$^{-1}$~\AA$^{-1}$. The photometry is measured in circular apertures with 3~pixels (0$^{\prime\prime}$.12) radius on the plane of the sky. Aperture corrections and corrections for the attenuation due to the foreground dust from the Milky Way have been applied to the listed photometry.s See text for more details.}
\tablenotetext{}{(13) The logarithm of the luminosity in the Pa$\beta$ emission line, in units of erg~s$^{-1}$.}
\tablenotetext{}{(14) The logarithm of the equivalent width (EW) of Pa$\beta$, in \AA, calculated from the ratio of the emission line flux  to the stellar continuum flux density. }
\end{deluxetable}

\bibliographystyle{aasjournal}
\bibliography{bibliography_ngc4449}{}

\end{document}